# Coincidental Correctness in the Defects4J Benchmark

*Rawad Abou Assi, Chadi Trad, Marwan Maalouf, and Wes Masri**
American University of Beirut
Electrical and Computer Engineering Department
Beirut, Lebanon 1107 2020
*{ria21, cht02, mjm30, wm13}@aub.edu.lb*
**Corresponding author*

## Abstract

*Coincidental correctness* (*CC*) arises when a defective program produces the correct output despite the fact that the defect within was exercised. Researchers have recognized the negative impact of coincidental correctness, and the authors have previously conducted a study demonstrating its prevalence in test suites. However, that study was limited to system tests and small subjects seeded with artificial defects. In this paper, we conduct a wider scope study of *CC* that addresses the following research questions in the context of the *Defects4J* benchmark:

**RQ1:** *Is CC prevalent in Defects4J?*
**RQ2:** *Is CC affected by the testing levels in Defects4J?*
**RQ3:** *Do CC tests induce peculiar infection paths in Defects4J?*

Furthermore, we use *JTidy* and *NanoXML* to address the following question:

**RQ4:** *Are the infections likely to be nullified within or outside the buggy method?*

To answer RQ1, we manually injected two code checkers for each of the 395 *Defects4J* defects: 1) a *weak checker* that detects *weak CC* tests by monitoring whether the defect was reached; and 2) a *strong checker* that detects *strong CC* tests by monitoring whether the defect was reached and the program has transitioned into an infectious state. Our results showed that *CC* is prevalent in *Defects4J*, as we observed 38.1× more *strong CC* tests than failing tests and 60.5× more *weak CC* tests than failing tests.

Testing has traditionally been classified into several levels that include *unit*, *module*, *integration*, *system*, and *acceptance*. Meanwhile, the test cases in *Defects4J* are not classified into any of the aforementioned testing levels. In addition, the boundaries between such levels are not clear due to the lack of a clear universal definition. Therefore, in order to answer RQ2, we derive the testing level of a test case from its method coverage information; specifically, we base it on the number and frequency of execution of the methods it covers. Our results showed that *CC* is present at all testing levels, but is more prevalent in high testing levels than in low testing levels.

To answer RQ3, we contrasted the characteristics of the infection propagation paths induced by the *Defects4J* failing tests to those induced by the *strong CC* tests. We observed that the paths induced by the *CC* tests: 1) were considerably longer on average; and 2) comprised a higher number of conditional, modulo, multiplication, division, and invocation statements.

Finally, to answer RQ4 which relates to RQ2, we performed an experiment involving *JTidy*, *NanoXML,* and their associated high-level test suites. We used code checkers to determine whether, in the case of *strong CC*, the infections were nullified before exiting the buggy function or afterward. All of our observations showed that the infections were nullified after exiting the buggy function.

# 1. Introduction

*Coincidental correctness* (*CC*) [13][42][55][44] arises when a defective program produces the correct output despite the fact that the defect within was exercised. The *CC* phenomenon might seem appealing since it yields "good" behavior out of "bad" programs. However, there are two main problems introduced by coincidental correctness: 1) it results in overestimating the reliability of programs, since it hides defects that might subsequently surface following unrelated code modifications; and 2) it reduces the effectiveness of various program analysis techniques that aim at enhancing program reliability. Therefore, there is a need to study *CC* in order to learn what causes it and the circumstances under which it is more likely to occur.

Voas [55] presented a technique that predicts whether testing is likely to reveal a given fault. The technique is centered on the following three notions that collectively form the *PIE* model: fault **e**xecution, program state **i**nfection, and infection **p**ropagation to the output. Subsequently, Amman and Offutt [3] presented *RIP*, a model similar to the *PIE* model and also aims at capturing the relationship between faults and failures. *RIP* requires the following sequence of three conditions to be satisfied in order for a failure to be observed:

- *Reachability*$_{cond}$: the location (or locations) in the program that contain the fault must be reached.
- *Infection*$_{cond}$: the program's state must become incorrect, right after the fault is reached.
- *Propagation*$_{cond}$: The infected state must propagate to cause some output of the program to be incorrect.

In the context of the *RIP* model, a test case could be classified as follows [44]:

- *Failing*: if all three conditions were satisfied.
- *True Passing*: if none of the conditions were satisfied.
- *Weak CC*: if the *Reachability*$_{cond}$ was satisfied but none of the other two conditions were satisfied.
- *Strong CC*: if the *Reachability*$_{cond}$ and the *Infection*$_{cond}$ were satisfied but the *Propagation*$_{cond}$ was not satisfied.

As noted above, in this paper we consider two forms of *CC*, namely, *weak CC* and *strong CC*, which were previously proposed by two of the authors [43][44][45]. Keeping in mind that in the literature *CC* typically refers to *strong CC*.

Researchers have previously recognized the negative impact of *CC* on the effectiveness of coverage-based fault localization (CBFL) [10][57][44][22][9], model checking based fault localization [7], and testing [24][27]. The authors have also conducted a study demonstrating that both *weak CC* and *strong CC* are prevalent [43][44][45]. However, that study was mainly based on the *Siemens* benchmark (*sir.unl.edu*) [23], which comprises only system tests and programs that are no longer considered adequate by the testing community due their small sizes and the fact that they are seeded with artificial defects.

Meanwhile, the practice of unit testing has grown tremendously in recent years due to the fact that it is mandated by widely adopted software development methodologies and practices, such as *Extreme Programming* (*XP*) [11], *Agile Development, Test-Driven Development* (*TDD*) [12], *Continuous Delivery* (*CD*) [16], and *DevOps* [16]. Typically in these environments, developers use frameworks such as *JUnit* (*junit.org*) and TestNG (*testing.org*) to write unit tests that are continuously updated as new failing behaviors are uncovered.

The previous study by Masri *et al* [43][44][45] did not involve sizable code with real defects nor unit tests, and was limited in scope. In this paper, we conduct a wider scope study of *CC* that addresses the following research questions in the context of the *Defects4J* benchmark [32]:

**RQ1:** *Is CC prevalent in Defects4J?*

**RQ2:** *Is CC affected by the testing levels in Defects4J?*

**RQ3:** *Do CC tests induce peculiar infection paths in Defects4J?*

Furthermore, we use *JTidy* and *NanoXML* to address the following question:

**RQ4:** *Are the infections likely to be nullified within or outside the buggy function?*

To answer RQ1, we manually injected two code checkers for each of the 395 *Defects4J* defects: 1) a *weak checker* that detects *weak CC* by monitoring whether the $Reachability_{cond}$ was satisfied; and 2) a *strong checker* that detects *strong CC* by monitoring whether the $Reachability_{cond}$ and the $Infection_{cond}$ are both satisfied. Our results showed that *CC* is prevalent in *Defects4J*, as we observed 38.1× more *strong CC* tests than failing tests and 60.5× more *weak CC* tests than failing tests.

Testing has traditionally been classified into several levels that include *unit*, *module*, *integration*, *system*, and *acceptance*. In the meantime, the test cases in *Defects4J* are not classified into any of the aforementioned testing levels. In addition, there is no strong agreement amongst researchers and practitioners regarding the boundaries between such levels, which caused some practitioners to define their own levels and nomenclature. For example, Google uses only three levels, namely, *small*, *medium*, and *large* [58]. These three testing levels vary in terms of how many functions or modules they cover, on their time of execution, and on whether they could access file systems, networks, or databases (as opposed to relying on mocks and faked environments). Therefore, in order to answer RQ2, we derive the testing level of a test case from its method coverage information; specifically, we base it on the number and frequency of execution of the methods it covers. Our results showed that *CC* is present at all testing levels, but is more prevalent in high testing levels than in low testing levels.

To answer RQ3, we contrasted the characteristics of the infection propagation paths induced by the *Defects4J* failing tests to those induced by the *strong CC* tests. The aim is to identify any correlation between the $Propagation_{cond}$ not getting satisfied (*strong CC*) and the execution frequency of specific types of statements. We observed that the paths induced by the *CC* tests: 1) were considerably longer on average; and 2) comprised a higher number of conditional, modulo, multiplication, division, and invocation statements (see Section 2.3).

To answer RQ4 (which relates to RQ2), we performed an experiment involving multiple defective versions of *JTidy*, *NanoXML,* and their associated high-level test suites. We used code checkers to determine whether, in the case of *strong CC*, the infections were nullified before exiting the buggy function or afterward. Specifically, for each version we specified a first *strong checker* right after the defect and a second *checker* right before the return statement(s) of the function containing the defect. A passing test that always triggers the first checker but not the second is deemed to be a *CC* test in the context of low-level testing. A passing test that always triggers both checkers is deemed to be a *CC* test in the context of high-level testing. All of our observations showed that the infections were nullified after exiting the buggy function.

The main contributions of this work are:

- Experimental studies suggesting the following:
  - *CC* is prevalent in sizable code comprising real defects.
  - *CC* is prevalent across testing levels, but much more so on the high levels.
  - Compared to failing tests, *CC* tests induce infection propagation paths that are likely to be longer, and comprise a higher number of conditional, modulo, multiplication, division, and invocation statements.
  - Infections are more likely to be nullified outside of the buggy function than within.
- An extension of the *Defects4J* benchmark encompassing a weak checker and a strong checker for each of its 395 defects. Researchers using *Defects4J* will therefore be able to factor out the negative effect of *CC* by treating passing tests that triggered our checkers as failing, or by discarding them altogether. This artifact is downloadable from [67].
- An extension of the *Defects4J* benchmark associating each test case with a (relative) testing level [67].

Section 2 presents background related to the prevalence of *CC* in system test suites. Sections 3 through 6 answer RQ1 through RQ4, respectively. Section 7 discusses threats to the validity of our experiments. Section 8 surveys related work, and Section 9 concludes.

## 2. Background

*Coincidental Correctness* has been studied by several researchers for various purposes [5][9][13][19][20][24][27][28][33][42][55][57]. However, our previous work [45][44][43] is most relevant, which we summarize in this section.

### 2.1 Prevalence of CC

In order to identify the factors that impair CBFL, Masri *et al* [45] conducted an empirical study that assesses the prevalence of several scenarios including *weak CC* and *strong CC*. Their study involved 148 seeded versions of ten Java programs that included seven programs from the *Siemens* benchmark (132 versions) and three releases of *NanoXML* (*sir.unl.edu*). The test suite sizes for the ten programs ranged from 140 to 4130, with a total number of 19873 test cases. Their study showed the following:

1) Out of the 19873 test cases, 3120 were *strong CC* and 11208 were *weak CC*, i.e., 15.7% and 56.4%, respectively.
2) 20 versions had more than 60% of their tests as *strong CC,* and 86 versions had more than 60% of their tests as *weak CC.*
3) 41 versions had none of their tests as *strong CC*, and only 5 versions had none of their tests as *weak CC.*
4) Interestingly, one version had 3133 out of its 3155 tests as *strong CC*, i.e., 99.3%.

The findings by Masri *et al* [45] are very interesting; however, the fact that they mostly involved the *Siemens* programs motivated us to further study *CC* using a more accepted benchmark, namely, *Defects4J.*

### 2.2 Impact of Weak CC on Coverage Techniques

Coincidental *Correctness*, in its *weak* form, is potentially detrimental to most coverage-based and test-based program analysis techniques, including fault localization [1][6][31][38][39][41][52], test-based program repair, test suite reduction, and test case prioritization [15][34][40][53][54][60]. Masri *et al* [45] (and later Masri and Abou Assi [44]) studied the negative impact of *weak CC* on coverage-based fault localization (CBFL). Specifically, they demonstrated that *weak CC* has a safety reducing effect on CBFL by analytically showing how the presence of *weak CC* tests leads to suspiciousness metrics that underestimate the suspiciousness of the faulty code. The metrics they analyzed included that of *Jaccard* [14], *Tarantula* [31], *AMPLE* [22], and *Ochiai* [1].

We now illustrate by example how *weak CC* can have a negative impact on CBFL, and on test suite reduction and test case prioritization. The code shown in Table 1 is widely used in the literature [31], it is meant to compute the median of three input numbers. Line 6 is defective as it assigns `y` to `m` as opposed to assigning it `x`. Table 1 also shows six test cases and their corresponding statement coverage information: a check mark indicates that the statement at the given row was executed at least once using the test case at the given column. As indicated in the table, $t_6$ is the only failing test case. And as expected, $t_6$ executes the defect at Line 6. Meanwhile, $t_1$ also executes the defect at Line 6 but outputs the correct result, which makes it a *CC* test case. Specifically, $t_1$ is a *weak* and not a *strong CC* test case since it executes the defect but does not cause an infection at Line 6. It erroneously assigns y to m as opposed to assigning x, but since x and y are both 3, the program state does not get infected.

| `int median(int x, int y, int z) {` | Test Cases | | | | | |
|---|---|---|---|---|---|---|
| | $t_1$ | $t_2$ | $t_3$ | $t_4$ | $t_5$ | $t_6$ |
| | 3, 3, 5 | 1, 2, 3 | 3, 2, 1 | 5, 5, 5 | 5, 3, 4 | 2, 1, 3 |
| `1: int m = z;` | ✓ | ✓ | ✓ | ✓ | ✓ | ✓ |
| `2: if (y < z)` | ✓ | ✓ | ✓ | ✓ | ✓ | ✓ |
| `3:     if (x < y)` | ✓ | ✓ | | | ✓ | ✓ |
| `4:         m = y;` | | ✓ | | | | |
| `5:     else if (x<z)` | ✓ | | | | ✓ | ✓ |
| `6:         m = y;   // bug: m = x;` | ✓ | | | | | ✓ |
| `7: else` | | | ✓ | ✓ | | |
| `8:     if (x > y)` | | | ✓ | ✓ | | |
| `9:         m = y;` | | | ✓ | | | |
| `10:    else if (x>z)` | | | | ✓ | | |
| `11:        m = x;` | | | | | | |
| `12: return m;` `}` | P | P | P | P | P | F |

**Table 1.** Java code and corresponding statement coverage information

The fact that $t_1$ is a *CC* test diminishes the correlation between the execution of Line 6 and failure, which translates into lessening the effectiveness of CBFL. In other words, including $t_1$ in the test suite yields a suspiciousness score at Line 6 that is lower than the score computed when $t_1$ is excluded. For example, the value of the *Tarantula* suspiciousness score [31] for Line 6 is 0.833 when $t_1$ is included and 1.0 when excluded.

A widely adopted test suite reduction (TSR) approach discards redundant tests while insuring that the program elements covered by the original test suite are also covered by the reduced test suite. In our example, the test suite would be reduced to $T_1 = \{t_1, t_2, t_3, t_4\}$ or to $T_2 = \{t_2, t_3, t_4, t_6\}$ since these are the only minimal test suites that cover all statements covered by the original test suite. Consequently, there will only be a 50% chance that the reduced test suite will reveal the defect since: 1) $T_1$ and $T_2$ are equally likely to be generated (using greedy TSR [34][48]); and 2) only $T_2$ includes the failing test $t_6$. Excluding the *weak CC* test $t_1$ from the original test suite would mean that only $T_2$ will be generated by test suite reduction; a clearly more desirable outcome.

A classical test case prioritization approach gives higher execution priority to the test cases that cover the most not yet covered program elements. A large number of prioritization outcomes are possible for our example, of which we list only two that represent extreme cases, namely, $T_1 = <t_1, t_3, t_4, t_2, t_5, t_6>$ and $T_2 = <t_6, t_3, t_4, t_2, t_5, t_1>$. Clearly, $T_2$ is superior to $T_1$, since the defect is revealed by the first executed test in $T_2$ as opposed to the last executed test in $T_1$. Here also, excluding $t_1$ from the original test suite would result in a more effective prioritization. Actually, in case $t_1$ is excluded, $t_6$ will always be executed either first or second.

### 2.3 Causes of Strong CC

In the case of *strong CC*, the $Infection_{cond}$ is satisfied but the $Propagation_{cond}$ is not. This could be interpreted as information loss along the path starting at the infection location and ending at the output. Voas and Miller [56] devised a metric that approximates the information loss between the input and output of a mathematical function or a program statement (or block of statements). They termed the metric *domain-to-range ratio* or DRR, which is the ratio of the cardinality of the possible inputs to the cardinality of the possible outputs; a larger DRR represents higher information loss.

Masri and Abou Assi [44] illustrated the use of DRR using the three snippets of code below:

```
S1: y = x * 3;

S2: y = x % 3;

S3: if (x >= 3) {
        y = 1;
    } else {
        y = 0;
    }
```

For each snippet, assume that the input variable x takes on the values [1, 5], of which the value 4 represents an infection.

Computing DRR for $S_1$ we get DRR($S_1$) = 5/5 = 1, which means that $S_1$ will not cause information loss, and thus will enable the infection to propagate without any obstruction.

Considering $S_2$, DRR($S_2$) = 5/3 = 1.67, which means that $S_2$ might prevent the infection from propagating. Actually, when x is infected (i.e., x is 4), y takes on the value 1. However, y being 1 does not represent a propagated infection since y will also be 1 when x is not infected (i.e., x is 1).

Similarly, DRR($S_3$) = 5/2 = 2.5, indicating that $S_3$ might also prevent the infection to propagate. Specifically, y becomes 1 when x is 3 or 4 or 5, i.e., when x is infected and when it is not infected. Therefore, y being 1 does not represent a propagated infection.

**Buggy Code**

```
res[rOff] =  (tmp2[0] <= 0) ? -PI : PI) - 2 * tmp2[0];

for (int i = 1; i < tmp2.length; ++i)    {
      res[rOff + i] = -2 * tmp2[i];
}
```

**← missing statement**

**Fixed Code**

```
res[rOff] =  (tmp2[0] <= 0) ? -PI : PI) - 2 * tmp2[0];

for (int i = 1; i < tmp2.length; ++i)  {
      res[rOff + i] = -2 * tmp2[i];
}

// Fix
res[rOff] = atan2(y[yOff], x[xOff]);
```

**Buggy Code + Checkers**

```
res[rOff] =  (tmp2[0] <= 0) ? -PI : PI) - 2 * tmp2[0];

for (int i = 1; i < tmp2.length; ++i)    {
      res[rOff + i] = -2 * tmp2[i];
}
println("Weak Checker #10");


if (res[rOff] != atan2(y[yOff], x[xOff])) {
         println("Strong Checker #10");
}
```

**Figure 1** – Simple *Infection*$_{cond}$ checker *(Math, Bug #10)*

**Buggy Code**

```
if (str == null || searchStr == null) {
   return false;
}
result = contains(str.toUpperCase(),
searchStr.toUpperCase());
return result;
```

**Fixed Code**

```
if (str == null || searchStr == null) {
   return false;
}

// Fix
int len = searchStr.length();
int max = str.length() - len;
for (int i = 0; i <= max; i++) {
    if (str.regionMatches(true, i, searchStr, 0, len)) {
       return true;
   }
}

return false;
```

**Buggy Code + Checkers**

```
if (str == null || searchStr == null) {
   return false;
}
result = contains(str.toUpperCase(),
searchStr.toUpperCase());

println("Weak Checker #40");

boolean fixedResult = false;
int len = searchStr.length();
int max = str.length() - len;
for (int i = 0; i <= max; i++) {
   if (str.regionMatches(true, i, searchStr, 0, len)) {
      fixedResult = true;
      break;
   }
}
if (result != fixedResult) {
   println("\nStrong Checker #40");
}
return result;
```

**Figure 2** – Non-trivial *Infection*$_{cond}$ checker *(Lang, Bug #40)*

It goes without saying that code constructs similar to $S_2$ and (especially) $S_3$ are pervasive, which explains the non-negligible presence of *strong CC* in the field. Such presence is also supported by the results presented by Androutsopoulos *et al* [5], which showed that 10% of the test cases involved in their experiments were *strong CC* tests.

Finally, in relation to information loss, Masri and Podgurski [47] conducted a study that revealed that most dynamic dependences in programs do not convey any measurable information. Such finding, which corroborates the fact that

| *Name* | *#v* | *\|T(v)\|* | | | *∑\|T(v)\|* | *\|$T_{bug}$(v)\|* | | | *∑\|$T_{bug}$(v)\|* |
|---|---|---|---|---|---|---|---|---|---|
| | | **min** | **max** | **avg** | | **min** | **max** | **avg** | |
| ***Chart*** | 26 | 1,586 | 2,193 | 1,814 | 47,244 | 1 | 223 | 24 | 622 |
| ***Closure*** | 133 | 2,595 | 8,444 | 7,203 | 958,339 | 2 | 6324 | 812 | 106,398 |
| ***Lang*** | 65 | 1,437 | 2,291 | 1,817 | 114,314 | 1 | 111 | 13 | 821 |
| ***Math*** | 106 | 818 | 4,378 | 2,512 | 262,503 | 1 | 434 | 30 | 2,290 |
| ***Mockito*** | 38 | 838 | 1,395 | 1,251 | 42,710 | 5 | 917 | 210 | 6,524 |
| ***Time*** | 27 | 3,749 | 4,041 | 3,916 | 105,723 | 2 | 736 | 118 | 3,190 |

**Table 2** – *Defects4J* Libraries

*$Propagation_{cond}$* is not always satisfied, could also be explained by the frequent occurrence of constructs similar to $S_2$ and $S_3$.

## 3. RQ1: *Is CC Prevalent in Defects4J?*

This section aims at assessing the presence of coincidental correctness in a large codebase comprising real defects. We opted to use the *Defects4J* benchmark [32] for the following reasons:

1) It involves a large number of real defects with corresponding fixes.
2) It includes a large number of *JUnit* tests with oracles.
3) It is currently the de-facto benchmark for major research areas including software testing, automated fault localization, and (recently) automated program repair.
4) The outcome of our study will serve as an extension to *Defects4J* that will enable researchers to factor out the effect of coincidental correctness during their analyses.

Our primary goal here is to classify the test cases in *Defects4J* as *Failing*, *True Passing*, *Weak CC*, or *Strong CC*. For that, we need the means to determine which of the three *RIP* conditions were satisfied during a given test run. Our approach relies on the following three entities being available for each defect in our subject programs: 1) the location(s) of the bug; 2) the bug fix; and 3) the test oracle. Which are all readily available in *Defects4J*.

The satisfiability of the *$Reachability_{cond}$* is checked by simply injecting the buggy version with a print statement(s) at the location(s) of the bug, a *weak checker*. The satisfiability of the *$Propagation_{cond}$* is checked by the test oracle. The satisfiability of the *$Infection_{cond}$* is programmatically determined by injecting a code checker near the bug, a *strong checker*. This code checker is (manually) inferred by comparing the buggy code and the bug fix, and can vary in complexity from a simple conditional to numerous lines of code. Figure 1 shows an example of a simple *strong checker*, whereas Figure 2 shows a somewhat complex *strong checker*; both written to help identify *CC* tests in the *Defects4J* benchmark. It is worth noting that the nature of some bugs makes them inherently immune to coincidental correctness. For example, a bug due to a Java class having a missing

"implements Serializable" could never induce a *CC* test, since any test written to verify the serialization functionality of the class will always fail. No code checkers need to be injected when dealing with bugs of such nature.

The task of specifying the code checkers for the 395 bugs in *Defects4J* was carried out in several phases: 1) 24 undergraduate students in an advanced programming class were each assigned 10 bugs; 2) 11 students were selected out of the 24 based on the quality of the strong checkers they delivered, and were assigned between 30 to 40 bugs (which included the initial 10 they already worked on); 3) each of the 395 checkers were independently validated by two of the authors.

Table 2 describes the six libraries encompassing *Defects4J*, namely, *JFreeChart* (*Chart*), *Closure*, *Apache commons-lang* (*Lang*), *Apache commons-math* (*Math*), *Mockito*, and *Joda-Time* (*Time*). A number of versions of each library are provided for a total of 395 defective versions (see column 2). Each version *v* contains a single defect and is associated with its own test suite $T(v)$. Columns 3 to 6 in Table 2 show statistics about $T(v)$, namely, the min, max, average of $|T(v)|$ and the sum of $|T(v)|$ across all versions, i.e., $\sum|T(v)|$. Note that we observed that test suites associated with different versions might partially contain the same test cases.

Columns 7 to 10 in Table 2 also show statistics about $T_{bug}(v)$, which is the subset of test cases in $T(v)$ that directly or indirectly invoke the method containing the defect (but not necessarily reach the defect). Considering $T_{bug}(v)$ stems from the fact that some tests are meant to exercise a very specific functionality, therefore, many tests might not be relevant to the defect at hand. $T_{bug}(v)$ excludes such tests. Apparently, based on the considerable differences between $\sum|T(v)|$ and $\sum|T_{bug}(v)|$, most tests associated with a given version are not meant to exercise functionalities that are relevant to the defect.

Figure 3 shows the results of the test cases breakdown for *Chart, Lang, Math, Mockito, Time*, and *Closure*. Figure 4 shows breakdown for the libraries combined. Each plot shows the cumulative results from all the versions for a given library, specifically: 1) $\sum|Fail(v)|$: the number of failures observed across all versions; 2) $\sum|strongCC(v)|$: the number of *strong CC* tests detected across all versions; 3) $\sum|weakCC(v)|$: the number of *weak CC* tests detected across all versions; 4) $\sum|truePass(v)|_{bug}$: the number of true passing tests in all versions that belonged to $T_{bug}(v)$ (i.e., tests that entered the method containing the defect but did not reach the defect); and 5) $\sum|truePass(v)|$: the number of true passing tests in all versions that belonged to $T( v)$. More importantly, the plots are annotated with the ratios $\sum|strongCC(v)|/\sum|Fail(v)|$ and $\sum|weakCC(v)|/\sum|Fail(v)|$. These ratios are likely to provide a good predictive power on how harmful coincidental correctness will be on techniques such as fault localization, test suite reduction, and test-based program repair. In CBFL for example, as the value of either $\sum|strongCC(v)|/\sum|Fail(v)|$ or $\sum|weakCC(v)|/\sum|Fail(v)|$ increases, the value of the suspiciousness metric associated with the defect decreases. Also, in coverage-based test suite reduction, an increase of either ratios will likely yield reduced test suites that might not contain the failing test cases. *Hereafter, we will use* $\sum|strongCC(v)|/\sum|Fail(v)|$ *as a*

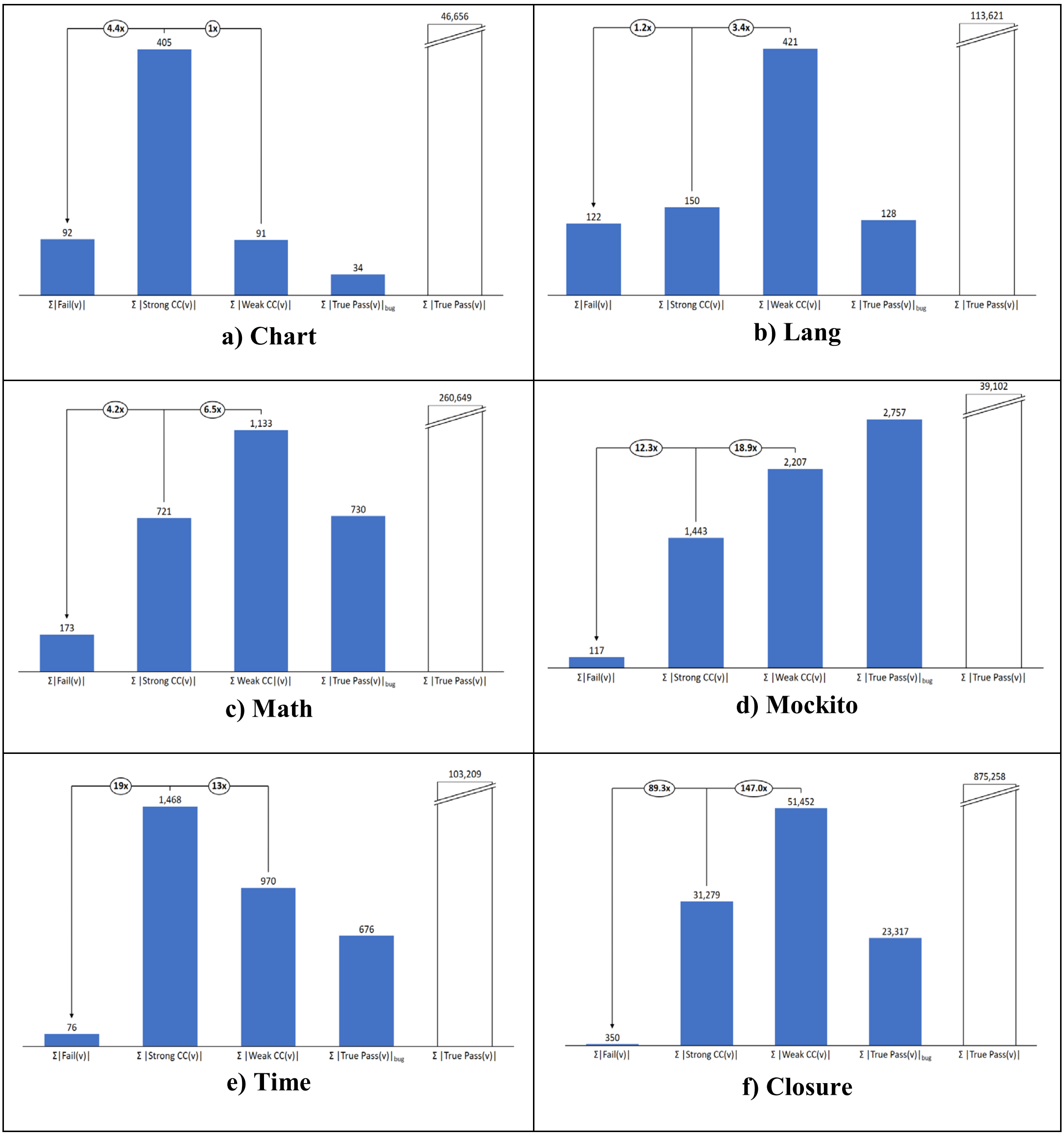

**Figure 3** – Test case classification

*measure of the prevalence of strong CC, and ∑|weakCC(v)|/∑|Fail(v)| as a measure of the prevalence of weak CC.*

As an example, Figure 3.e shows that executing the 27 test suites of *Time* on their respective defective versions involved: 1) 76 failures; 2) 1,468 strong *CC* tests; 3) 970 *weak CC* tests; 4) 676 true passing tests that are relevant to the defect; and 5) 103,209 true passing tests that might or might not be relevant to the defect. Figure 3.e also shows that there are 19 times more *strong CC* tests than failures, and 13 times more *weak CC* tests than failures.

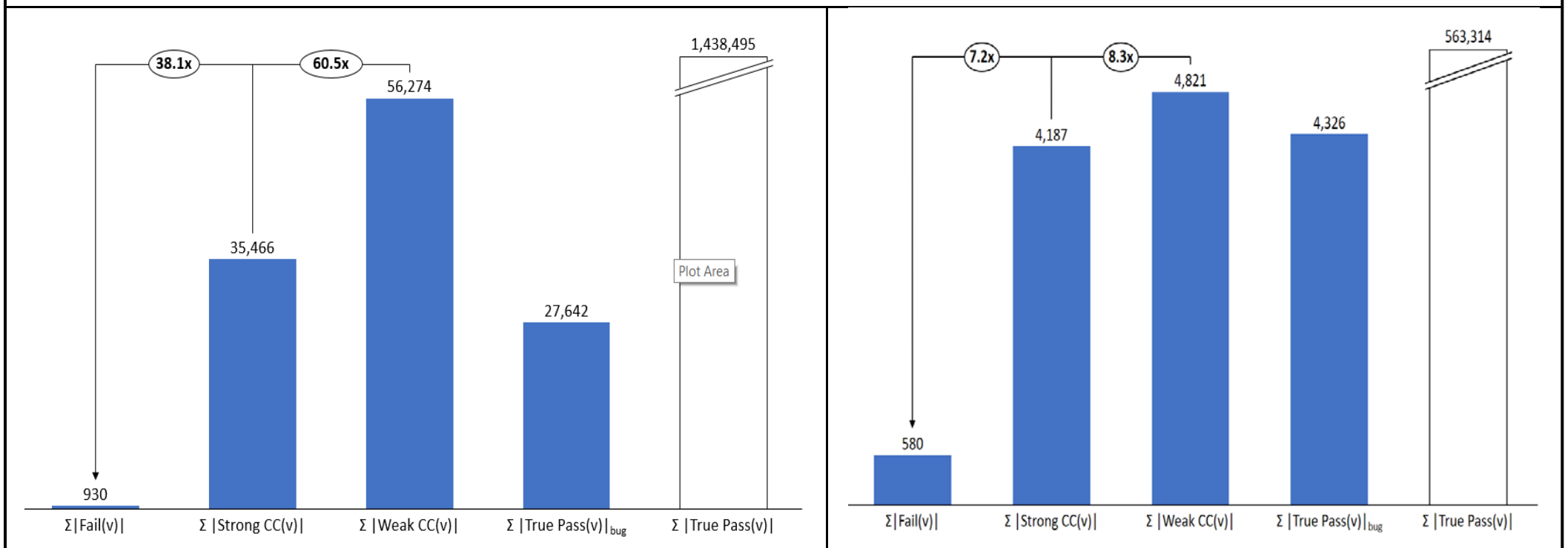

**a)** *Closure* included **b)** *Closure* excluded

**Figure 4 –** Test case classification for the libraries combined

A more interesting example is that of *Closure*. As shown in Figure 3.f, there are 89.3 times more *strong CC* tests than failures, and 147 times more *weak CC* tests than failures. This suggests that *CC* is more likely to have a negative impact on defect detection in *Closure* than in any of the other libraries.

For each of the six libraries, the number of combined *weak* and *strong CC* tests was considerably larger than the number of failures. In the case of *Time* and *Chart*, there were more *strong CC* tests than *weak CC* tests, which is somewhat unexpected. Considering all six libraries combined, Figure 4.a shows the test cases breakdown as follows: 1) 930 failures; 2) 35,466 *strong CC* tests; 3) 56,274 *weak CC* tests; 4) 27,642 true passing tests that belonged to $T_{bug}(v)$; and 5) 1,438,495 true passing tests that belonged to $T(v)$. The following statistics provide a sensible summary of Figure 4.a:

$$\sum|strongCC(v)|/\sum|Fail(v)| = 38.1$$
$$\sum|weakCC(v)|/\sum|Fail(v)| = 60.5$$
$$\sum|strongCC(v)|/\sum|T_{bug}(v)| = .30$$
$$\sum|weakCC(v)|/\sum|T_{bug}(v)| = .47$$
$$\sum|strongCC(v)|/\sum|T(v)| = .023$$
$$\sum|weakCC(v)|/\sum|T(v)| = .036$$

The ratios $\sum|strongCC(v)|/\sum|T(v)|$ and $\sum|weakCC(v)|/\sum|T(v)|$ seem relatively low; this is explained by the fact that the absolute majority of test cases do not invoke the functions containing the defects. On the other hand, the values exhibited by $\sum|strongCC(v)|/\sum|T_{bug}(v)|$ and $\sum|weakCC(v)|/\sum|T_{bug}(v)|$ are significant, which suggests that *CC* is prevalent amongst test cases that do invoke the functions containing the defects. Finally, as mentioned earlier, the ratios $\sum|strongCC(v)|/\sum|Fail(v)|$ and $\sum|weakCC(v)|/\sum|Fail(v)|$ are most relevant to predicting the negative impact of *CC*; and their values happened to be very significant.

Given that *Closure* is the predominant library with respect to the number of test cases, we opted to also consider all libraries combined, excluding *Closure*. Figure 4.b shows the resulting test case breakdown, and the summarizing statistics are as follows:

$$\sum|strongCC(v)|/\sum|Fail(v)| = 7.2$$
$$\sum|weakCC(v)|/\sum|Fail(v)| = 8.3$$
$$\sum|strongCC(v)|/\sum|T_{bug}(v)| = .31$$
$$\sum|weakCC(v)|/\sum|T_{bug}(v)| = .36$$
$$\sum|strongCC(v)|/\sum|T(v)| = .007$$
$$\sum|weakCC(v)|/\sum|T(v)| = .008$$

Excluding *Closure* still resulted in significant values of $\sum|strongCC(v)|/\sum|Fail(v)|$ and $\sum|weakCC(v)|/\sum|Fail(v)|$. Therefore, our answer to **RQ1** is:

**"Yes, *CC* is prevalent in *Defects4J*, which is sizable and comprises real defects"**

## 4. RQ2: *Is CC Affected by the Testing Levels in Defects4J?*

As stated earlier, there is no convergence in the testing community about the boundaries of the traditional testing levels, i.e., *unit*, *module*, *integration*, *system*, or *acceptance* [3][4]. Given a test targeting a specific method $m_1$ in class $c_1$, would that test still be classified as a *unit* test if it invokes a (helper or initialization) method $m_2$? What if $m_2$ is in a different class $c_2$? What if several methods other than $m_2$ are also involved? What if $m_1$ interacts with a database, or invokes a stored procedure in a database? The answers to such questions are very subjective. (We had an email exchange with the first author of *Defecst4J* [31] who supported this view.)

In this study, instead of trying to classify the test cases in *Defects4J* according to the traditional breakdown, we derive the testing level of a test case from its method coverage information. Basically, more coverage means higher testing level. First, we consider the number methods covered by the test case, then we consider the frequency of execution of such methods. In this manner, the testing level represents a (simple) relative measure that we can use as the basis for comparing tests within *Defects4J*.

***Testing levels based on the number of covered methods*** - Figure 5 shows for each library the boxplot of the exhibited testing levels based on the number of covered methods. For example, the mean testing level in *Closure* is 734 covered methods, in *Mockito* it is 173, in *Time* it is 69, in *Chart* it is 51, in *Math* it is 30, and in *Lang* it is 9 covered methods. Except for the case of *Closure*, the boxplots in Figure 5 do not clearly convey detailed information about the distribution of the testing levels. Figure 6 addresses this issue by presenting the boxplots using a logarithmic scale. Based on Figures 5 and 6, it is sensible to rank the six libraries in terms of the testing levels of their respective test suites as follows: *Lang* < *Math* < *Chart* < *Time* < *Mockito* < *Closure*. Note how the testing levels for *Closure* are considerably higher than for the other libraries. In fact, some researchers consider the tests provided for *Closure* to be *system* tests, whereas the others as *unit* tests. (This was pointed out by Jahangirova and Tonella in a discussion we had at CREST57 [20]). Figure 7 shows for each library a bar graph representing the prevalence of *strong CC*, *weak CC*, and *CC* (i.e., both forms combined). Specifically, the left bar represents $\sum|strongCC(v)|/\sum|Fail(v)|$, the middle bar represents $\sum|weakCC(v)|/\sum|Fail(v)|$, and the right bar represents their sum. By comparing the ratios in Figure 7 to the boxplots of Figures 5 and 6, the apparent trend (although not very strong) is that as the testing level increases the prevalence of *CC* increases. In addition, given all the test cases in all six libraries, Figure 8 shows the prevalence of coincidental correctness across different testing levels. Specifically, the observed testing levels are partitioned into ten intervals, and for each interval it shows the ratios $\sum|strongCC(v)|/\sum|Fail(v)|$, $\sum|weakCC(v)|/\sum|Fail(v)|$, and $(\sum|strongCC(v)|+ \sum|weakCC(v)|)/\sum|Fail(v)|$ in sequence. Here also, the apparent trend is that as the testing level increases, the prevalence of *CC* increases.

***Testing levels based on the frequency of execution of the covered methods*** – Figure 9 shows for each library the boxplot (using a logarithmic scale) of the exhibited testing levels based on the frequency of execution of the covered methods. Similar to Figures 5 and 6, the testing levels for *Closure* are considerably higher than for the other libraries. In addition, one could rank the libraries as follows in terms of testing levels: *Lang* < *Chart* < *Math* < *Time* < *Mockito* < *Closure*. This is similar to the previous ranking with the exception of *Chart* and *Math* being swapped. Furthermore, similar to Figure 8, Figure 10 shows the prevalence of coincidental correctness across different testing levels using all the tests from all the libraries. The trend in Figure 10 is comparable to that in Figure 8.

Finally, our observations largely converged based on either of our definitions of a *testing level*. Therefore, the answer to **RQ2** is:

> ***"Yes, CC*** **is affected by the testing levels in** ***Defects4J.*** **Specifically, it is present at all testing levels, and as the testing level increases it becomes more prevalent."**

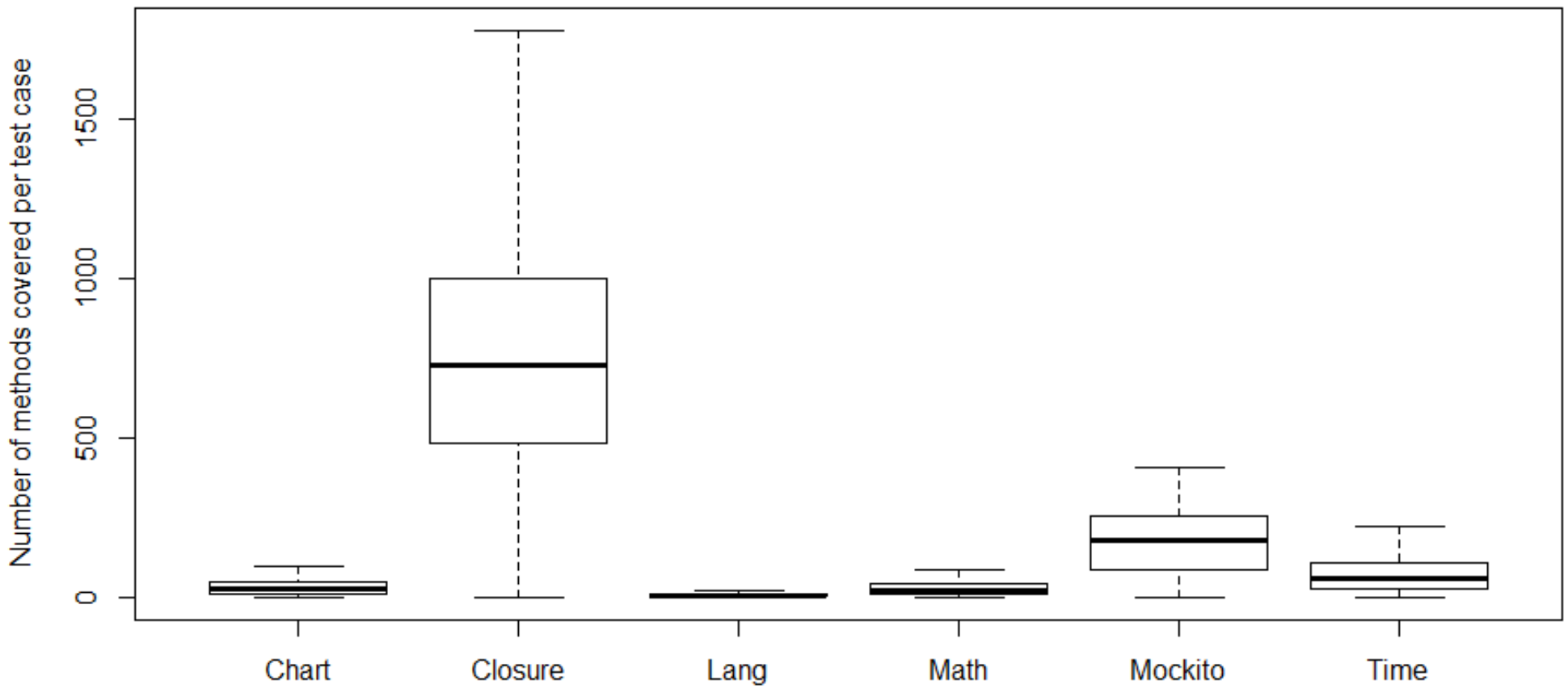

**Figure 5** – Distribution of testing levels per library

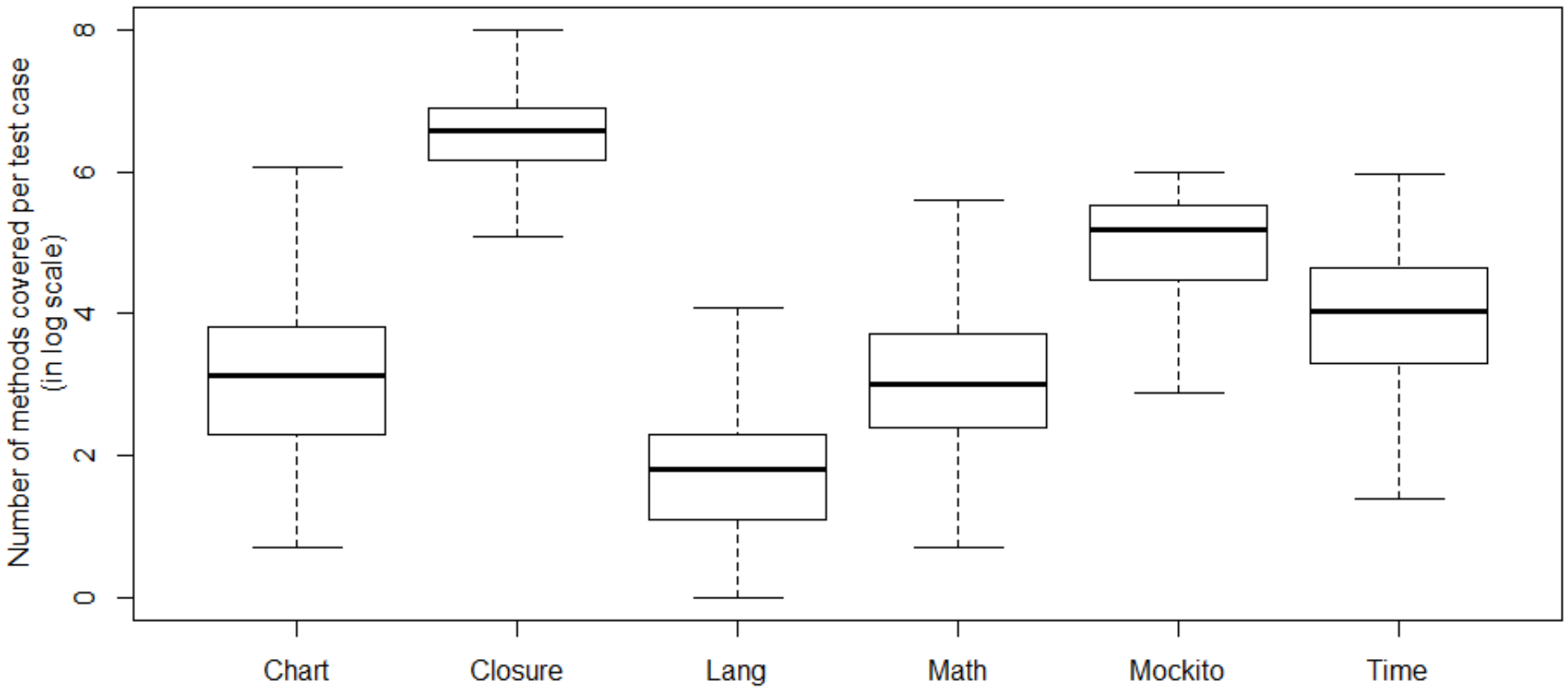

**Figure 6** – Distribution of testing levels per library (logarithmic scale)

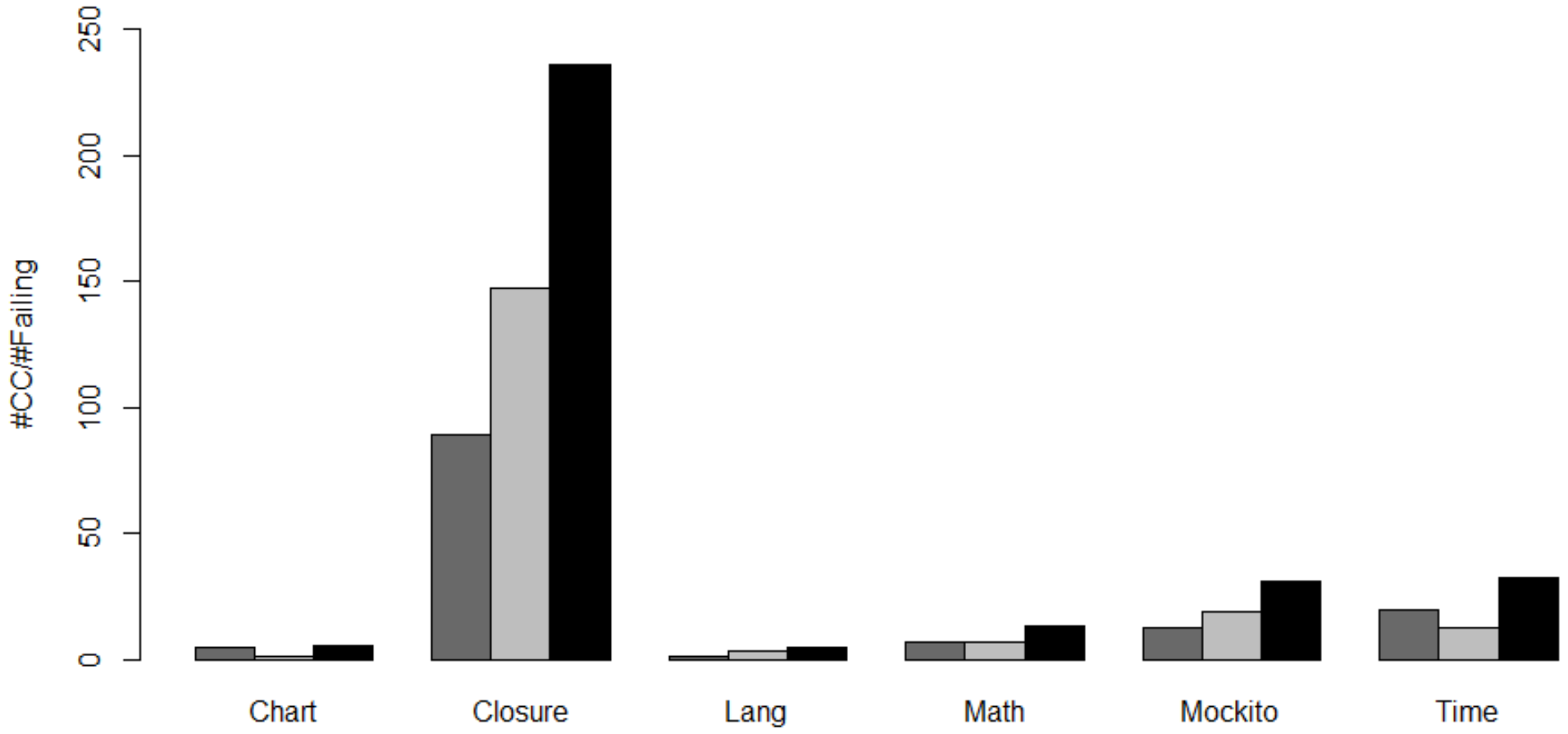

**Figure 7** – Prevalence of *strong CC*, *weak CC*, and *CC* per library

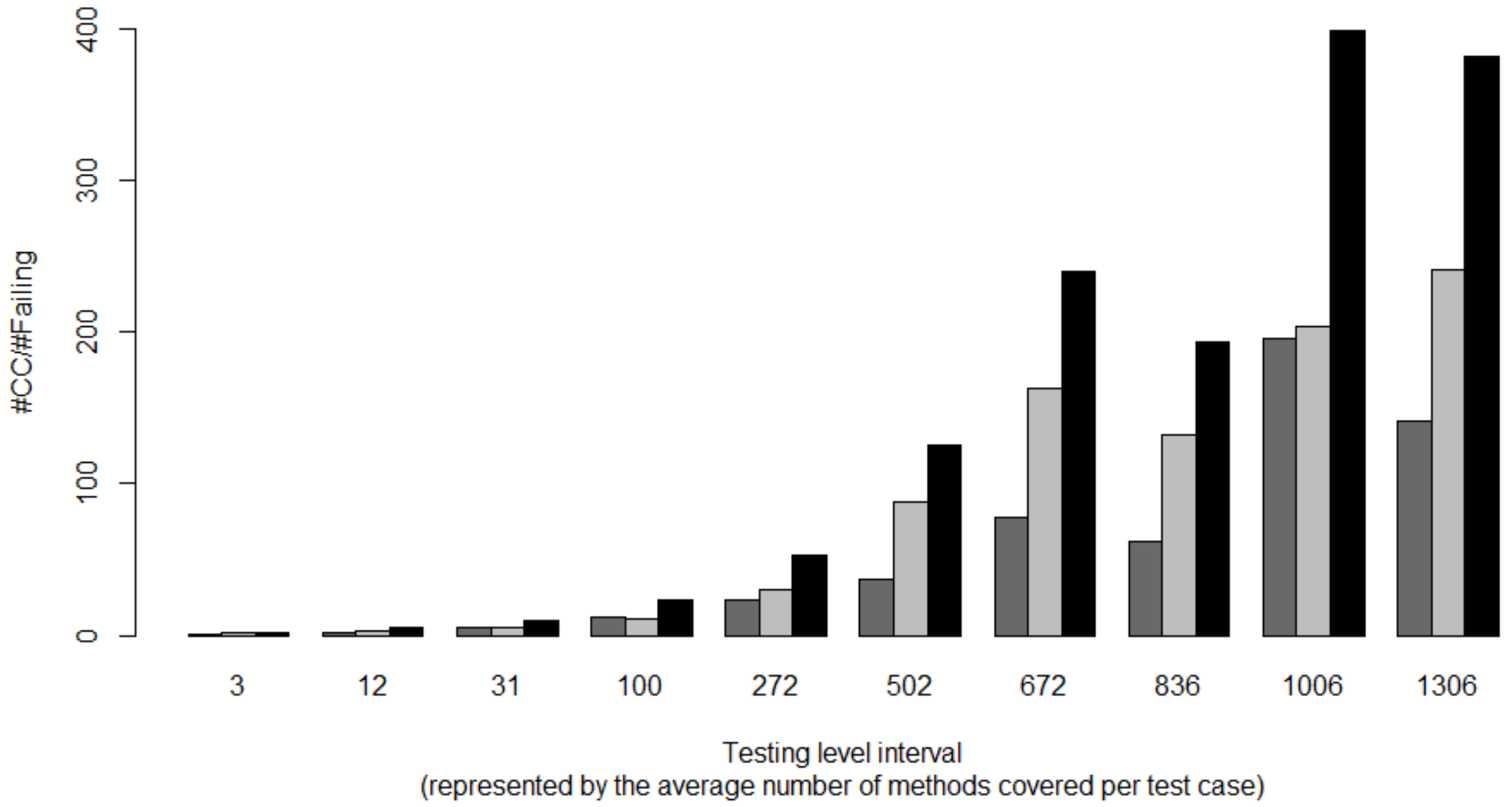

**Figure 8 –** Prevalence of *strong CC*, *weak CC*, and *CC* across testing level intervals

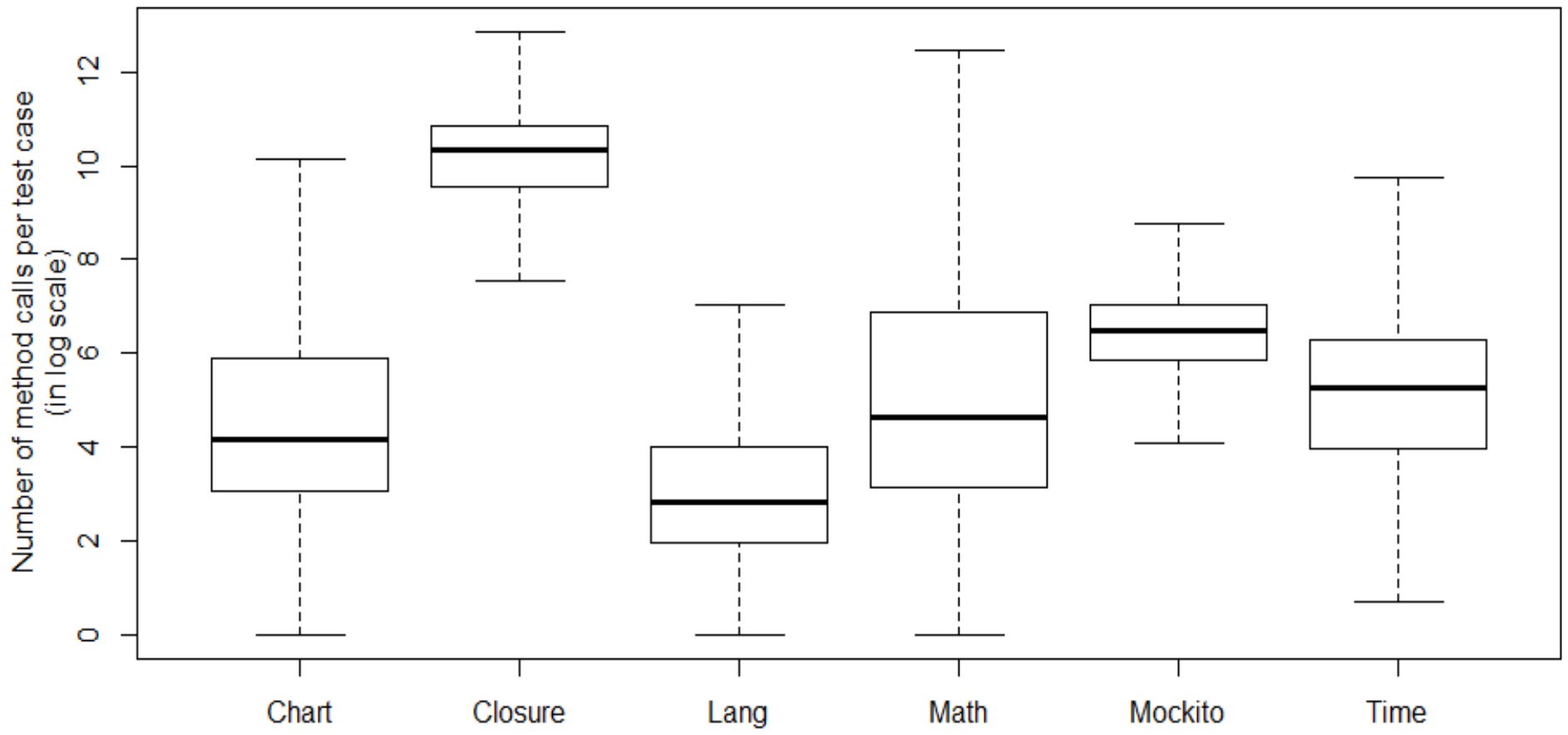

**Figure 9 –** Distribution of testing levels per library (logarithmic scale)

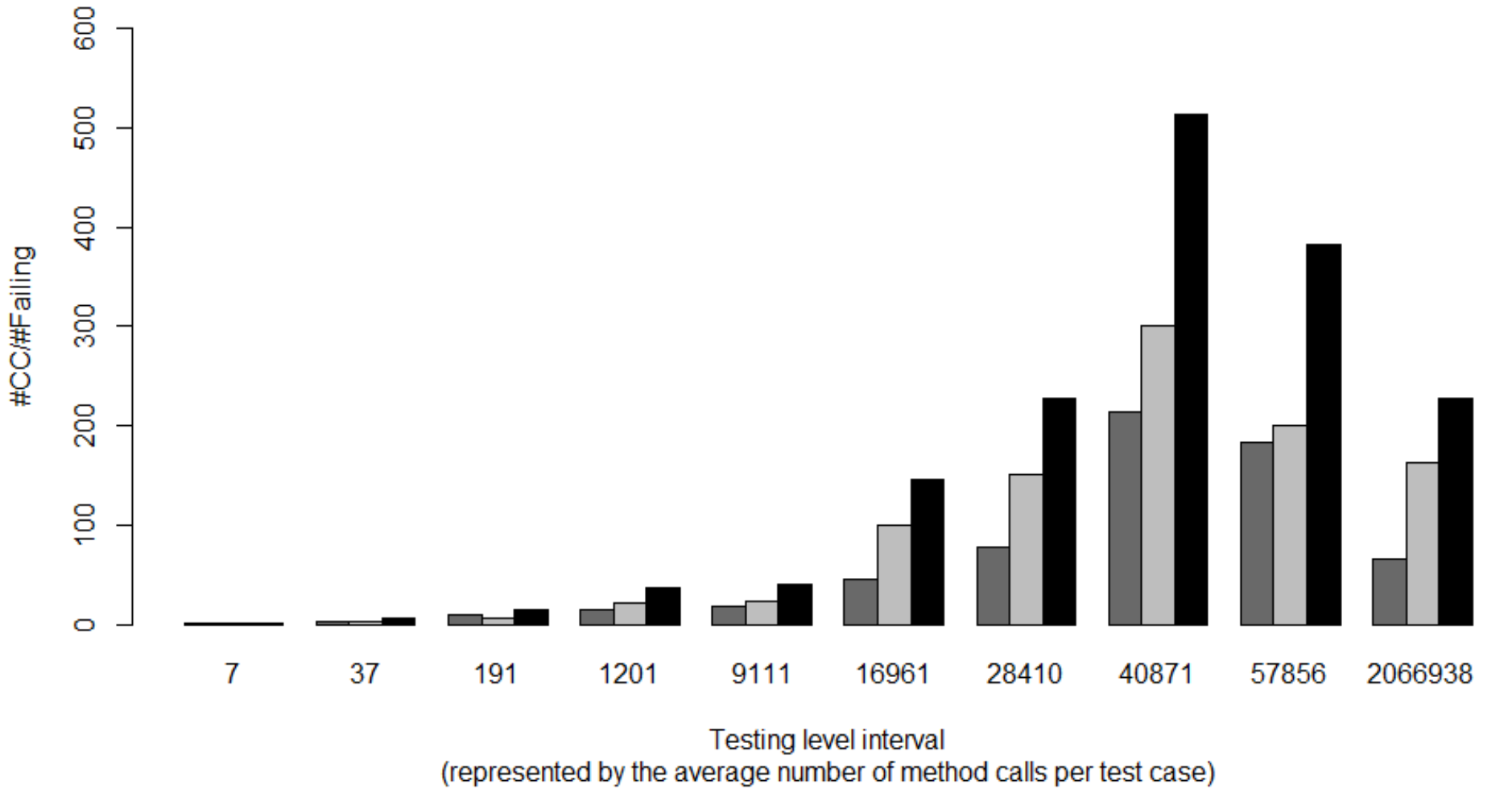

**Figure 10 –** Prevalence of *strong CC*, *weak CC*, and *CC* across testing level

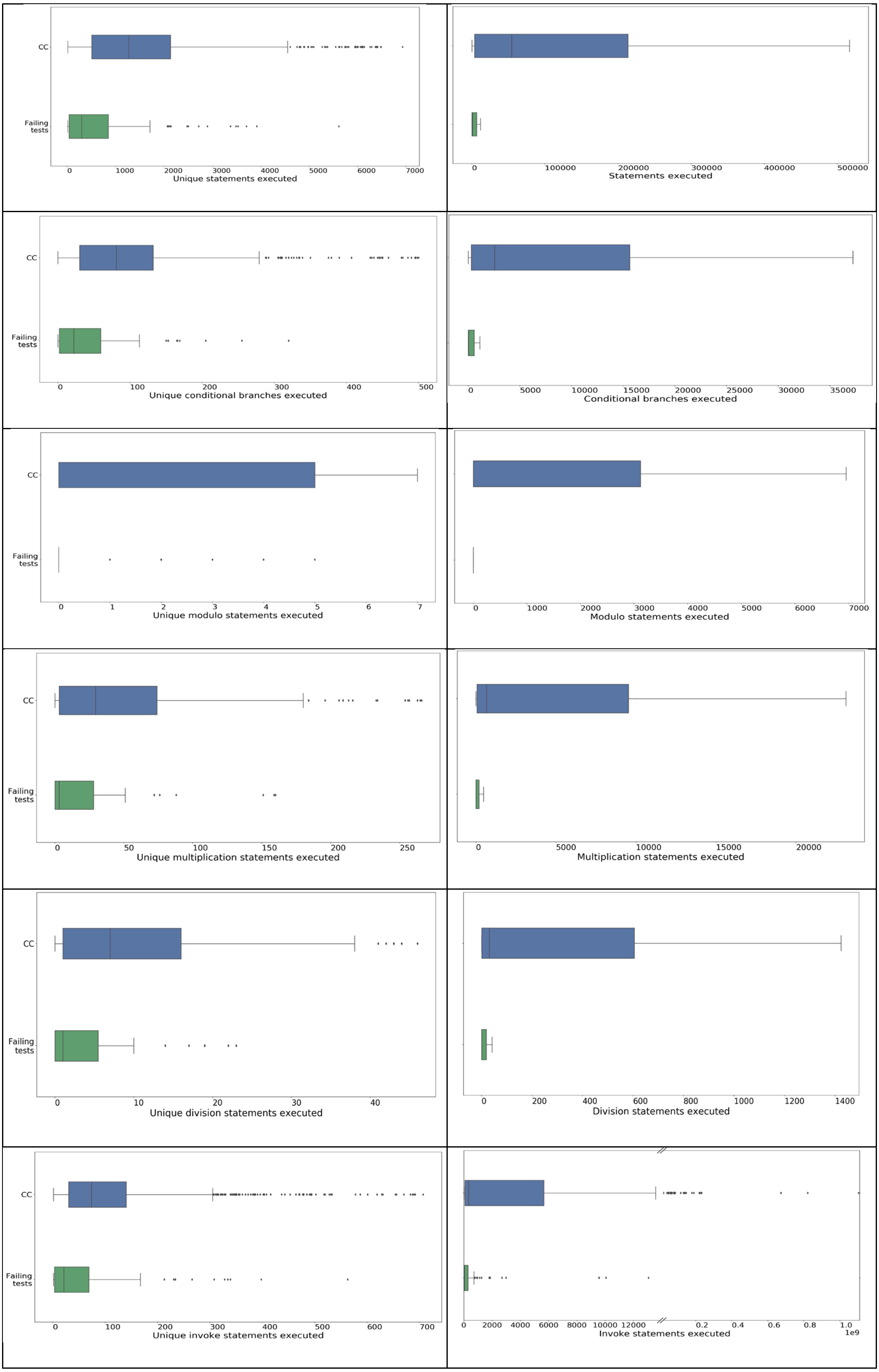

**Figure 11 –** Propagation analysis for *Math*

## 5. RQ3: *Do CC Tests Induce Peculiar Infection Paths in Defects4J?*

*Strong CC* is due to the $Propagation_{cond}$ not being satisfied. In Section 2.3 we attempted to understand this phenomenon by pointing out some simple programming scenarios that might cause it. This section sheds more light on the issue by empirically quantifying the execution frequency of program statements that are likely to nullify the propagation of an infectious state. This analysis, which was restricted to the six *Defecst4J* libraries, required us to build a profiler in order to collect the relevant data. The profiler, which targets the Java platform, was built based on the ASM Java bytecode manipulation and analysis framework (*asm.ow2.org*) and comprises *1.1K LOC.*

Our analysis was carried out as follows:

1) Once the $Infection_{cond}$ is satisfied, i.e., when a *strong checker* is triggered (right after the fault is reached), the profiler is initiated to start collecting the execution frequency of the following entities until the output is reached: a) all statements; b) conditional statements; c) modulo statements; d) multiplication statements; e) division statements; and f) invocation statements.
2) For each entity, two counters are updated, one that tracks its unique occurrences and another that tracks all its occurrences.
3) The collected profiles are categorized as belonging to failing tests or to *strong CC* tests.

Figure 11 shows the resulting boxplots for the *Math* library; the plots for the remaining libraries could be found in *Appendix-A*. In the plots, the upper whisker is bounded by (Q3 + 1.5 × IQR) and the lower whisker is bounded by (Q1 - 1.5 × IQR), where the interquartile range IQR = Q3 - Q1. Outliers are shown as individual points that fall outside the range confined by the upper and lower whiskers.

Each boxplot in Figure 11 shows the data collected as a result of executing the test suites associated with the 106 defective versions in *Math*. Specifically, there are 721 data points associated with the *strong CC* tests, and 173 data points associated with the failing tests (see Figure 3.c). Compared to the failing tests, the propagation paths induced by the *CC* tests in *Math* were considerably longer (on average) as they executed more conditional, modulo, multiplication, division, and invocation statements. It is worth noting that the differential is mostly apparent in the case of modulo statements. This overall trend is also exhibited by the other libraries but to varying levels, as shown in the plots shown in *Appendix-A*. Specifically, the trend is very strong in *Math*, *Chart*, *Mockito*, and *Lang*, but weak in *Closure* and *Time*.

The above findings are not surprising as a longer propagation path is more likely to encounter statements that can cause information loss, such as modulo and conditional statements. In addition, these results corroborate our findings in Section 4, i.e., *CC* tests are more likely to be prevalent in high testing levels.

Based on the results shown in Figure 11 and *Appendix-A*, the answer to **RQ3** is:

**"Compared to failing tests, *CC* tests were found to induce infection propagation paths that are more likely to be longer"**

## 6. RQ4: *Are the Infections Likely to be Nullified Within or Outside the Buggy Method?*

In this section, our focus will solely be on *strong CC*. For a more accurate comparison of the presence of *CC* in high-level testing as opposed to low-level testing, we performed the comparison using the same programs, bugs, and test cases. In this context, a high-level test is one that tries to exercise the application end-to-end by invoking `main()`, whereas a low-level test is one that exercises the buggy function $f_{bug}$.

We considered subject programs with existing high-level test suites that potentially include *CC* tests, namely, *NanoXML releases r1*, *r3*, and *r5*, and the *JTidy* HTML syntax checker and pretty printer *release 3* (subject programs that we used in previous work [43][44][45][47]). Given a *CC* high-level test, our aim is to determine whether or not an infection was nullified within the activation scope of the function that caused the infection, i.e., $f_{bug}$. We initially considered 16 single defect versions from *NanoXML* and 5 single defect versions from *JTidy*, but ended up discarding 7 *NanoXML* versions and 2 *JTidy* versions as they did not involve any *CC* tests. This left us with 4 versions from *NanoXML r1,* 5 versions from *NanoXML r5*, 3 versions from *JTidy*, and none from *NanoXML r3.* For the scope of this study, we believe that these numbers are adequate; in future work, we will target more versions and possibly complete benchmarks.

An overview of our analysis is shown in Figure 12. For each version, we specified three checkers: 1) $checker_{bug}$: inserted right after the defect; 2) $checker_{ret}$: inserted right before the return statement(s) of the function; and 3) $checker_{out}$: inserted at the program output to determine whether a test passed or failed. A passing test that triggers $checker_{bug}$ but not $checker_{ret}$ is deemed to be a *CC* test in the context of low-level testing; since the infection was nullified *within* the execution of $f_{bug}$. A passing test that triggers both $checker_{bug}$ and $checker_{ret}$ is deemed to be a *CC* test in the context high-level testing; since the infection was nullified *following* the execution of $f_{bug}$.

$checker_{bug}$ is inferred by comparing the buggy code to the fixed code, similar to the programmatic approach described in Section 3. $checker_{out}$ is based on a comparison of the output of the buggy version to that of the fixed version. $checker_{ret}$ is based on a comparison of the *partial program state* induced by the buggy function to that induced by the fixed function; a difference in states suggests that the infection has not been nullified when the buggy function returns. In our study, the *partial program state* induced by a buggy function $f_{bug}$ is represented by the *last* values assigned to variables between the time when $f_{bug}$ is entered and the time when it returns, regardless of whether or not the corresponding definition statements were located within $f_{bug}$. In addition, in case of a value-returning function, the return value is considered as part of the state. We opted to consider only the last assigned values since they collectively represent a snapshot of the state due to $f_{bug}$ that reaches the return statement. The above approach is similarly applied to the fixed function $f_{fixed}$.

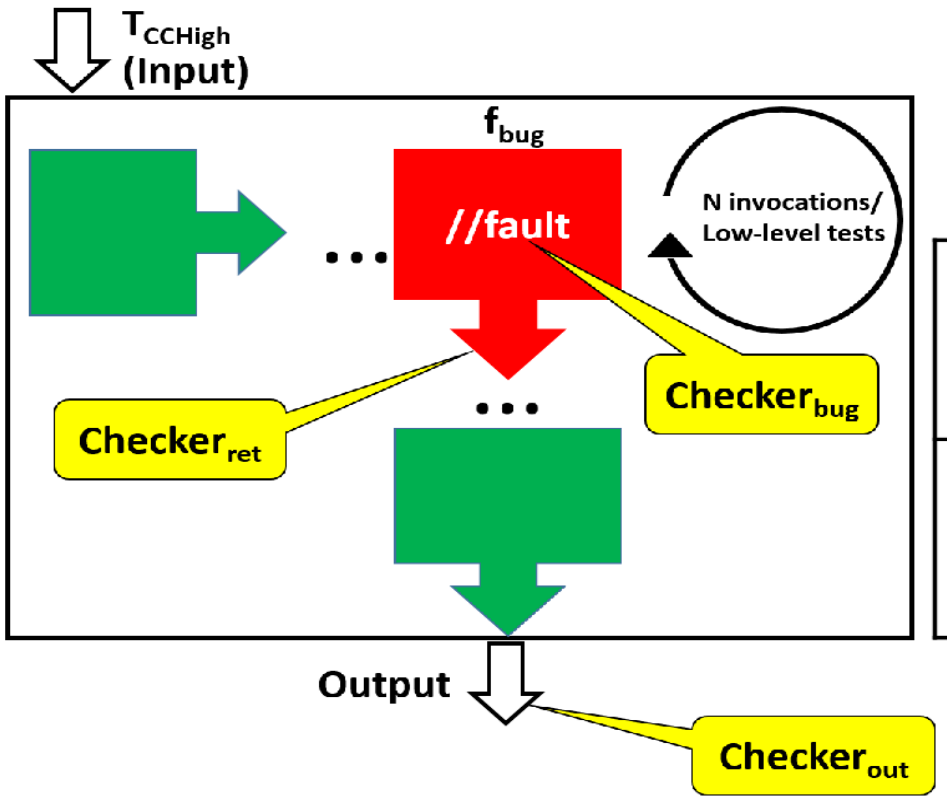

| | | $Checker_{bug}$ | $Checker_{ret}$ | $Checker_{out}$ |
|---|---|---|---|---|
| Low Level | Failing | ✓ | ✓ | |
| | True Passing | × | × | |
| | CC | ✓ | × | |
| High Level | Failing | ✓ | | ✓ |
| | True Passing | × | | × |
| | CC | ✓ | | × |

**Figure 12 –** Infection nullification Overview

A predominant behavior that we observed in our study is that the execution of a *CC* high-level test $t_{CCHigh}$ results in $f_{bug}$ getting invoked more than once, and $checker_{bug}$ getting triggered in some invocations but not in others. This behavior could be viewed as if a single $t_{CCHigh}$ induced $N$ low-level tests targeting $f_{bug}$. Each of the $N$ low-level tests could be classified as: a) *failing*: if both $checker_{bug}$ and $checker_{ret}$ were triggered; b) *true passing*: if neither $checker_{bug}$ nor $checker_{ret}$ were triggered; or c) *CC*: if $checker_{bug}$ was triggered but $checker_{ret}$ was not. We also recognize the following three categories of a given *CC* high-level test $t_{CCHigh}$:

1) ***IN***: $t_{CCHigh}$ **induced** $N$ **low-level tests that were only *CC* or *true passing*.** In this scenario, one could strongly assert that each infection was nullified with**in** the same activation scope of $f_{bug}$ that caused it. The pervasiveness of this scenario suggests that *strong CC* is more prevalent in low-level testing than in high-level testing.
2) ***OUT***: $t_{CCHigh}$ **induced** $N$ **low-level tests that were only *failing* or *true passing*.** Here, one could strongly assert that each infection was nullified **out**side the activation scope of $f_{bug}$ that caused it. The pervasiveness of this scenario suggests that strong *CC* is more prevalent in high-level testing than in low-level testing.
3) ***IN-OUT***: $t_{CCHigh}$ **induced** $N$ **low-level tests that were *CC*, *failing*, or *true passing*.** The pervasiveness of this scenario suggests two conflicting answers to RQ4 as some infections are nullified within and others outside the activation scope of $f_{bug}$ that caused them. For this reason, we will consider this scenario to be neutral in regard to RQ4.

In order to enable $checker_{ret}$, we built a profiler (*3.1K LOC*) also based on the ASM framework. The profiler records the values last assigned at return statements and definition statements involving: local variables, static variables, and instance field variables. While profiling a subject program, the profiler allows the user to start and stop the recording process, and to save the recorded states externally to a file. In our experiments, we manually injected calls to the profiler in $f_{bug}$ and $f_{fixed}$ requesting to start recording (at function entry), to stop recording (at function exits), and to externally save the collected states (also at function exits). This results in $N$ state files being created, one for each low-level test. Therefore, each $t_{CCHigh}$ will have two corresponding sets of state files, one set due to the $N$ invocations of $f_{bug}$ and another set due to $f_{fixed}$. Note that in our experiments both $f_{bug}$ and $f_{fixed}$ were always invoked the same number of times; i.e., $N$ times. The state files are then compared in pairs in order to determine the breakdown of the $N$ low-level tests. For example, the low-level test corresponding to the $i^{th}$ invocation of $f_{bug}$ is deemed *failing* if $checker_{bug}$ was triggered, and a mismatch is detected between the state file generated by the $i^{th}$ invocation of $f_{bug}$ and that generated by the $i^{th}$ invocation of $f_{fixed}$. One complicating factor that we (successfully) tackled when comparing state files relates to the fact that the same definition statements in $f_{bug}$ and $f_{fixed}$ might differ in terms of positioning; this was the case in two of the versions we studied, namely, $v_3$ of *JTidy* and $v_1$ of *NanoXML r1*. Also, given the current capabilities of our profiler, our present $checker_{ret}$ does not support defective functions that are recursive. Due to this shortcoming, we had to discard one version from *JTidy*, one from *NanoXML r1*, and one from *NanoXML r5*. As a result, our study was conducted using 3 versions from *NanoXML r1,* 4 versions from *NanoXML r5*, and 2 versions from *JTidy.*

The online supplement [67] provides the Java code for $f_{bug}$ and $f_{fixed}$ for each version, and for each $t_{CCHigh}$ within each version it provides: a) the number of invocations of $f_{bug}$ (i.e., $N$); b) the outcome of $checker_{bug}$ and $checker_{ret}$ for each of the $N$ invocations of $f_{bug}$; c) the state files for each of the $N$ invocations of $f_{bug}$ and those for $f_{fixed}$. Note that this paper only provides a very brief description of our profiler, which does not reflect its complexity. It is actually inspired by a profiler that we previously built for capturing state profiles [1].

The test suite for *JTidy* comprises 1000 tests (XML/HTML files), and the sizes of the test suites for the *NanoXML* releases are 214 for *r1*, 210 for *r3*, and 180 for

| | ***JTidy*** | | ***NanoXML r1*** | | | ***NanoXML r5*** | | | |
|---|---|---|---|---|---|---|---|---|---|
| | $v_2$ | $v_3$ | $v_1$ | $v_2$ | $v_3$ | $v_1$ | $v_2$ | $v_5$ | $v_6$ |
| $\|T_{CCHigh}\|$ | 8 | 15 | 31 | 22 | 36 | 176 | 57 | 70 | 49 |
| ***Avg N*** | 17.8 | 353.8 | 1 | 2.9 | 5 | 2 | 1.9 | 8.4 | 9.6 |
| $\|T_{CCHigh\text{-}IN}\|$ | 0 | 0 | 0 | 0 | 0 | 0 | 0 | 0 | 0 |
| $\|T_{CCHigh\text{-}OUT}\|$ | 8 | 15 | 31 | 22 | 36 | 176 | 57 | 70 | 49 |
| $\|T_{CCHigh\text{-}IN\text{-}OUT}\|$ | 0 | 0 | 0 | 0 | 0 | 0 | 0 | 0 | 0 |

**Table 3 –** Categorization of the *CC* high-level tests

*r5* [23]. Table 3 summarizes our results. For each version it shows $|T_{CCHigh}|$, the total number of *CC* high-level tests; the average number of invocations of $f_{bug}$; and the numbers of $t_{CCHigh}$ tests that belong to the *IN*, *OUT*, and *IN-OUT* categories, respectively. Note how the average value of *N* was greater than or equal to 2 in all except of two versions. More importantly, it is clear from Table 3 that category *OUT* is universal; i.e., for each version, 100% of the tests in $T_{CCHigh}$ belonged to $T_{CCHigh\text{-}OUT}$, and none belonged to $T_{CCHigh\text{-}IN\text{-}OUT}$ and $T_{CCHigh\text{-}IN}$. Note that if $checker_{ret}$ was not highly sensitive (or accurate), an undetected difference in states will result in a $t_{CCHigh}$ getting erroneously categorized under *IN* or *IN-OUT* instead of *OUT*. However, given that *IN* and *IN-OUT* are logically viable categories, we intend to conduct larger studies to demonstrate that fact.

Based on the results shown in Table 3, the answer to **RQ4** is:

> **"Based on a limited study, infections were found to be more likely to be nullified outside of the buggy function than within"**

## 6. Threats to Validity

A major threat to the external validity of our study is the fact that it was mainly focused on *Defects4J*, which is limited to only few domains and environments. Also, the oracles in *Defects4J* are not guaranteed to be free of flaws, and the test suites might include flaky tests. To address this issue, we will need to conduct further empirical studies using other benchmarks.

One threat to the internal validity of our approach is that for each bug, we implemented a *strong checker* in order to detect when the $Infection_{cond}$ is satisfied. Such critical checkers might not always be appropriately implemented especially in cases when the bug is due to deleted code or missing conditionals.

The checkers and oracles used in our experiments varied considerably as they were based on the following differing approaches: a) assertions placed at the output of *JUnit* tests (provided by *Defects4J*); b) conditional checks/specifications manually placed right after the defects; c) comparisons of program outputs; and d) comparisons of partial states of programs. One might argue that a single uniform approach should have been used throughout, for example, either a specification based approach or an approach based on comparing program states. However, writing accurate specifications that verify the correctness of functions might not be feasible especially for functions with considerable global side effects. On the other hand, accurately capturing and comparing complete program states is a very difficult endeavor. In fact, to our knowledge, all existing techniques for tracking and capturing program states adopt major approximation measures [30][64][65][66], which diminishes their accuracy.

Amman and Offutt [4] updated the *RIP* model into the *RIPR* model by refining the $Propagation_{cond}$ and adding the $Reveal_{cond}$ as follows:

- $Propagation_{cond}$: the infected state must cause some part of the final program state to become incorrect.
- $Reveal_{cond}$: the tester (or test oracle) must observe part of the incorrect final program state.

The updates highlight the fact that it is plausible that the final program state might be mostly infected, meanwhile, the test oracle might only consider the parts that are uninfected, thus, yielding a passing run. Adopting the more detailed *RIPR* model in our study in place of the *RIP*/*PIE* model is not necessary given our aims, as explained next. The *RIPR* model requires us to differentiate between two additional scenarios: a) the test case passes while the $Infection_{cond}$ is satisfied and the program state at the output is partially infected; and b) the test case passes while the $Infection_{cond}$ is satisfied and the program state at the output is not infected at all. The test case is deemed to be a *strong CC* test in both scenarios (since we rely on the oracle for correctness). Therefore, given that our main aim is to study coincidental correctness in *Defects4J*, and not necessarily validate the accuracy of its oracles, then there is no need to differentiate between these two scenarios. In other words, regardless of the model used (*RIP* or *RIPR*), our study would be similarly conducted and our conclusions would be identical.

## 7. Related Work

Section 2.1 presented our previous work that answers RQ1 in the context of the *Siemens* benchmark and *NanoXML*. Both studies converge by suggesting that both *weak CC* and *strong CC* are prevalent in test suites. However, the study in this paper is more valuable as it involves a larger codebase with real faults.

We are not aware of any previous work that studied coincidental correctness in relation to RQ2, RQ3, or RQ4. In this section we overview work involving *CC* from various perspectives, and when relevant, we provide a relative positioning of our current work.

***Early work*** –

The term *coincidental correctness* was first used by Timothy A. Budd in his PhD dissertation to refer to "mutations that produce the correct answer even though they are not correct" [13]. Marick [42] described *CC* as a problem that occurs whenever the weak mutation hypothesis (*WMH*) is not holding. *WMH* states that whenever a fault was executed and its effect is detectable at the fault location then the output will be affected. Laski *et al* [33] studied coincidental correctness while referring to it as *error masking*. They mutated the internal program state then checked whether or not the program produced the correct output. Their approach assesses the quality of a test suite and the presence of *CC* within. Voas [55] presented the *PIE* model which defines the three conditions that capture the relationship between faults and failures.

The above body of work provides the fundamentals behind coincidental correctness.

***Impact of CC on defect detection*** –

Forgacs and Bertolino [24] introduced the notion of *untested statements*, i.e., statements that failed to exhibit any influence on the output via dynamic dependence. They attributed this phenomenon to coincidental correctness. Ball *et al* [7] presented a fault localization technique that uses Model Checking to contrast the counterexamples (failing traces) to traces that conform to the property (passing traces), and reports the differences as fault localization aids to the developer. When evaluating their technique on 15 faults, they attributed its inability to locate three of the faults to coincidental correctness. Hierons [27] recognized the negative effect of *CC* when augmenting Partition Analysis with Boundary Value Analysis. He also showed how Boundary Value Analysis can be enhanced in order to reduce the likelihood of *CC* even in an environment that involves non-determinism and floating point numbers. Baudry *et al* [10] defined a dynamic basic block (DBB) as a set of statements that is covered by the same test cases. They empirically observed that test suites containing more DBB's resulted in improved CBFL. They also observed that the actual faulty DBB's were not always ranked as the most suspicious, which they attributed to *CC*. Masri *et al* demonstrated that *weak CC* has a safety reducing effect on CBFL by analytically showing how the presence of *weak CC* tests leads to suspiciousness metrics that underestimate the suspiciousness of the faulty code [44][43].

By recognizing the negative impact of *CC*, the aforementioned contributions motivated our work.

***Mitigation of CC*** –

Wang *et al* [57] introduced the notion of *context pattern*, which describes the data and control flow patterns that correlate with failure. In order to help mitigate the impact of *CC* on CBFL, they leveraged context patterns to refine code coverage to strengthen the correlation between program failures and the coverage of faulty statements. Two of the authors, Masri and Abou Assi [43][44] segregated test cases into two clusters based on their execution profiles; a passing test was deemed as *CC* if it fell within the cluster that contained most of the failing tests. They also presented a second technique in which only passing tests are partitioned into two clusters, conjecturing that the *CC* test cases will be grouped within the same cluster. The first technique presented by Masri and Abou Assi [44] was extended by several researchers in the aim of improving the quality of the clustering [49][45][35][36][37]. Also, similar to the first technique presented by Masri and Abou Assi [44], Bandyopadhyay and Ghosh [8][9] considered any passing test that is similar to failing tests as *CC*. More importantly, they proposed an iterative approach that leverages user feedback to improve the detection of *CC* tests and consequently CBFL. Masri *et al* [46] presented an interactive multivariate visualization based technique and tool that enables the user to identify *CC* tests. Similar to the work by Masri and Abou Assi [43][44], the approach conjectures that the execution profiles of *CC* tests are similar to those of failing tests.

Our study provides insight on the circumstances under which *CC* is more likely to occur, which could be beneficial to researchers working on *CC* mitigation.

***Information theory and CC*** -

Clark and Hierons [20] leveraged information theory to study *strong coincidental correctness*, which they termed *fault masking*. They derived an information theoretic measure, termed *squeeziness*, which quantifies the likelihood of a function *f* to nullify the propagation of an infectious state. The *squeeziness* of function $f\colon I \rightarrow O$ is $S_q(f) = H(I) - H(O)$, which is the loss of information after applying *f* to *I*. Their proposed measure somewhat relates to mutual information and the measure presented by Masri and Podgurski [47] for quantifying the amount of information flowing between two variables connected by a dynamic dependence chain. Clark and Hierons [20] also demonstrated that there is a strong statistical correlation between *squeeziness* and fault masking, whereas the correlation between DRR and fault masking was not as strong. Androutsopoulos *et al* [5] provided a more thorough information theoretic analysis of *strong CC,* which they termed *Failed Error Propagation* or *FEP*. They devised five metrics that aim at predicting the occurrence of *strong CC*, of which two showed a high predictive power. The results of the experiments they conducted corroborated our results and those presented by Masri *et al* [45], as 10% of the 7,140,000 involved test cases were *strong CC* tests. In a position paper, Clark *et al* [19] proposed information theory as the basis for solving several software engineering problems including the mitigation of coincidental correctness to increase the testability of programs.

Our current study does not treat *CC* from an information theoretic perspective. However, in a previous study [47], using an information theoretic measure, we showed that most dynamic dependences in programs do not convey any measurable information. Clearly, such finding relates to the prevalence of *strong CC*.

***Defects4J and CC*** -

In a Doctoral Symposium paper, Jahangirova [28] conjectured that it is beneficial to place oracles internally to a method as opposed to externally as this will decrease the likelihood of *strong CC* (or *FEP*). Jahangirova's aim is to study where to best place the oracles so as to maximize their overall fault detection ability. The placements to be explored are: a) externally, outside the method under test; b) at the return points of methods; and c) internal to the method. The plan is to use the defects in *Defects4J*'s while discarding the provided test suites that will be replaced by tests generated by EvoSuite [25], in order to better cover the faulty statements. The test oracles provided in *Defects4J* will also be discarded; instead, correctness will be determined by comparing the correct programs' states to the incorrect programs' states.

Our respective goals overlap in that we both need to measure the presence of *CC*; however, the approach of how it is done and the purpose of why it is done are very different. Jahangirova aims at determining where to best place test oracles, and we aim at: a) refining the *Defects4J* benchmark by identifying the *CC* tests within to be discarded if need be; and b) comparing the prevalence of *CC* at varying testing levels.

## 8. Conclusions

The results of our study showed that coincidental correctness is present at all testing levels in realistic code, and significantly more at the high testing levels. These results are not overly surprising but somewhat disappointing since they suggest that the negative impact of *CC* also holds at the low testing levels. Especially that the practice of unit testing (a low testing level) has grown tremendously in recent years.

A benefit that our work provides to researchers using the *Defects4J* benchmark is the ability to factor out the effect of coincidental correctness during their analyses, which will allow for more accurate evaluations of new reliability enhancing techniques.

Our study provided some evidence that *CC* is more likely to occur when the infection propagation path comprises a high number of conditional, and modulo statements. It also showed that infections are more likely to be nullified outside of the buggy function than within.

**Acknowledgment**

This research was supported in part by the Lebanese National Council for Scientific Research, and by the University Research Board at the American University of Beirut.

# Appendix-A

The plots below are referenced in Section 5. For each of *Chart, Lang, Mockito*, *Closure*, and *Time*, they present data similar to that of *Math* shown in Figure 11.

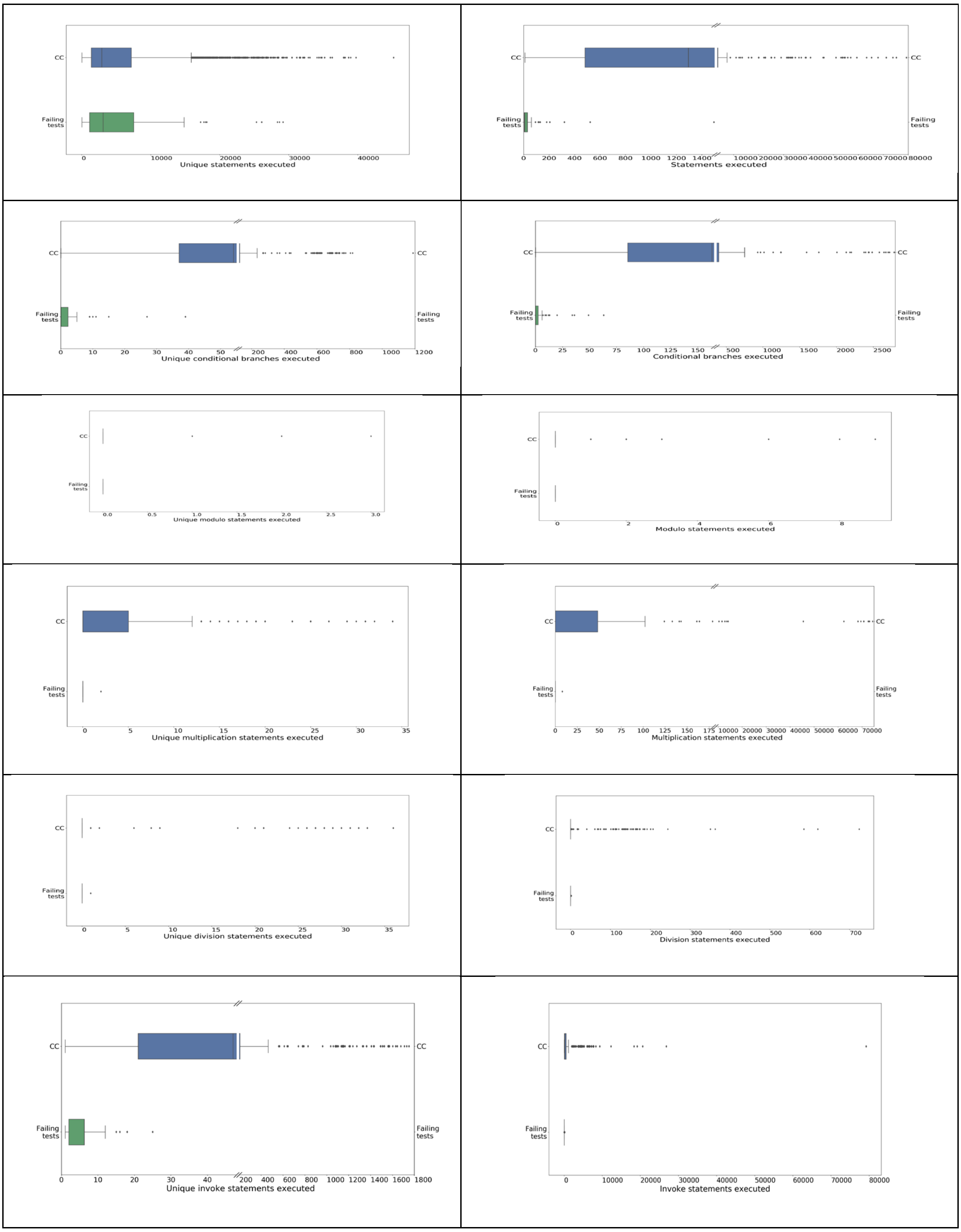

Propagation analysis for *Chart*

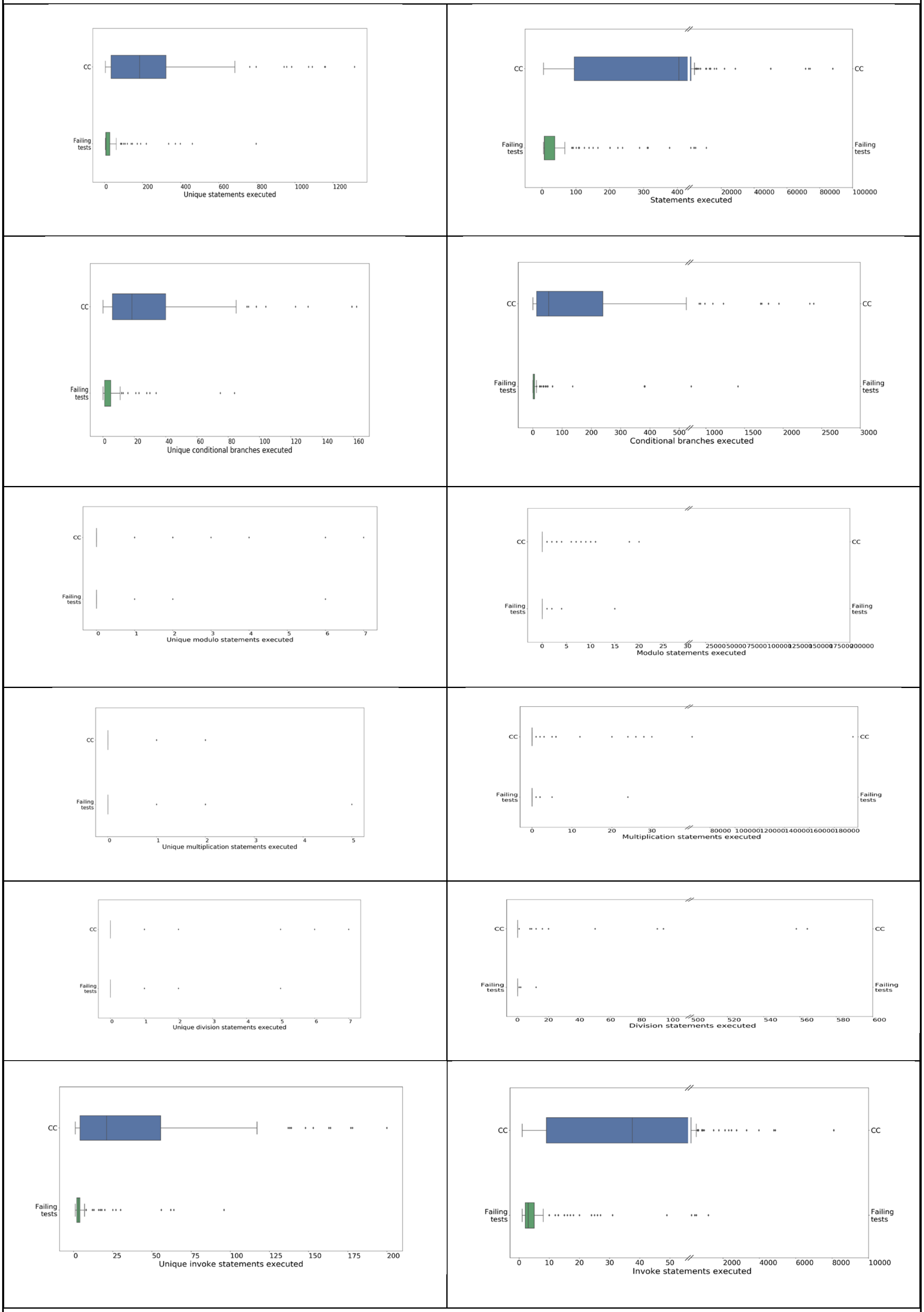

Propagation analysis for *Lang*

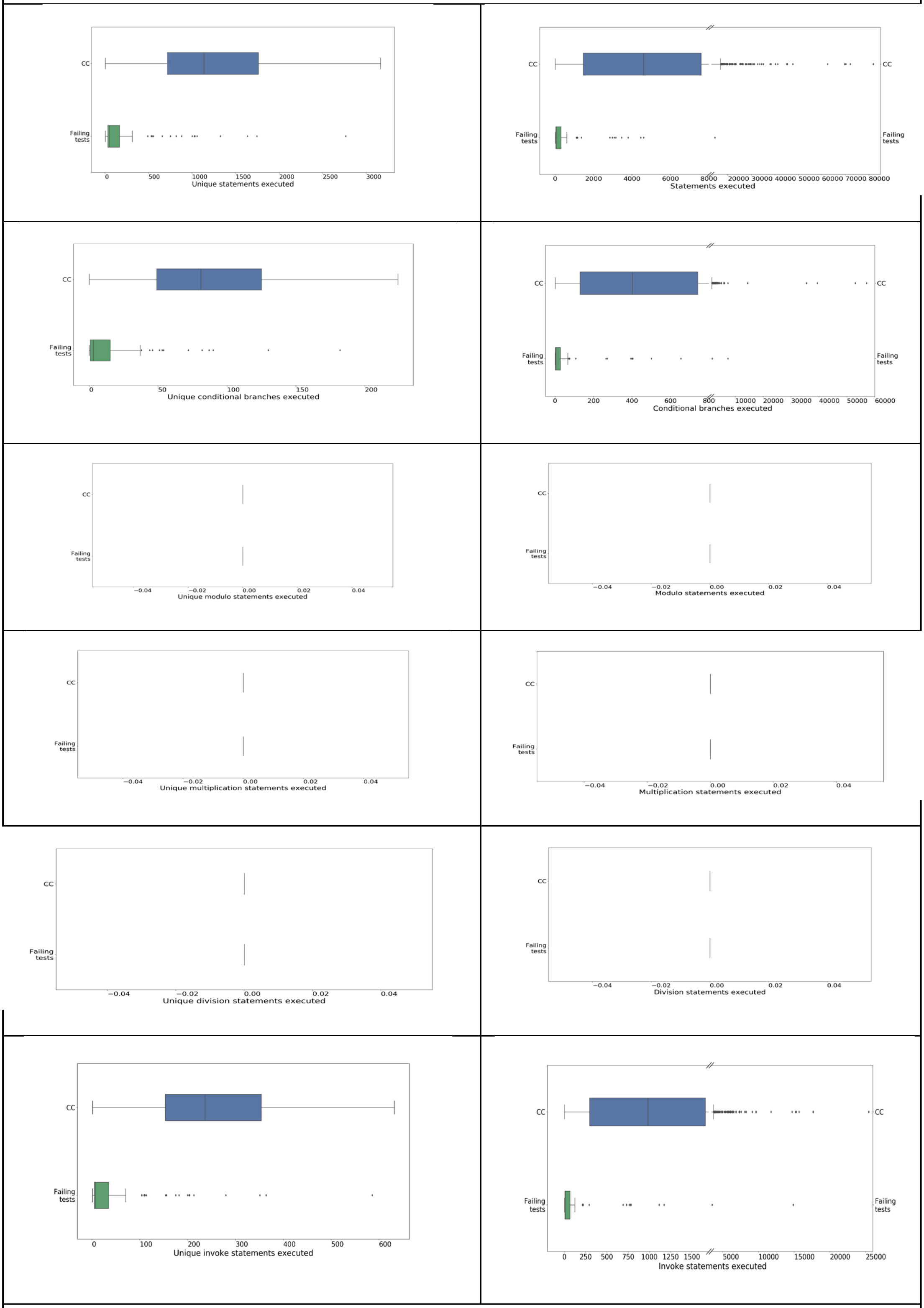

Propagation analysis for *Mockito*

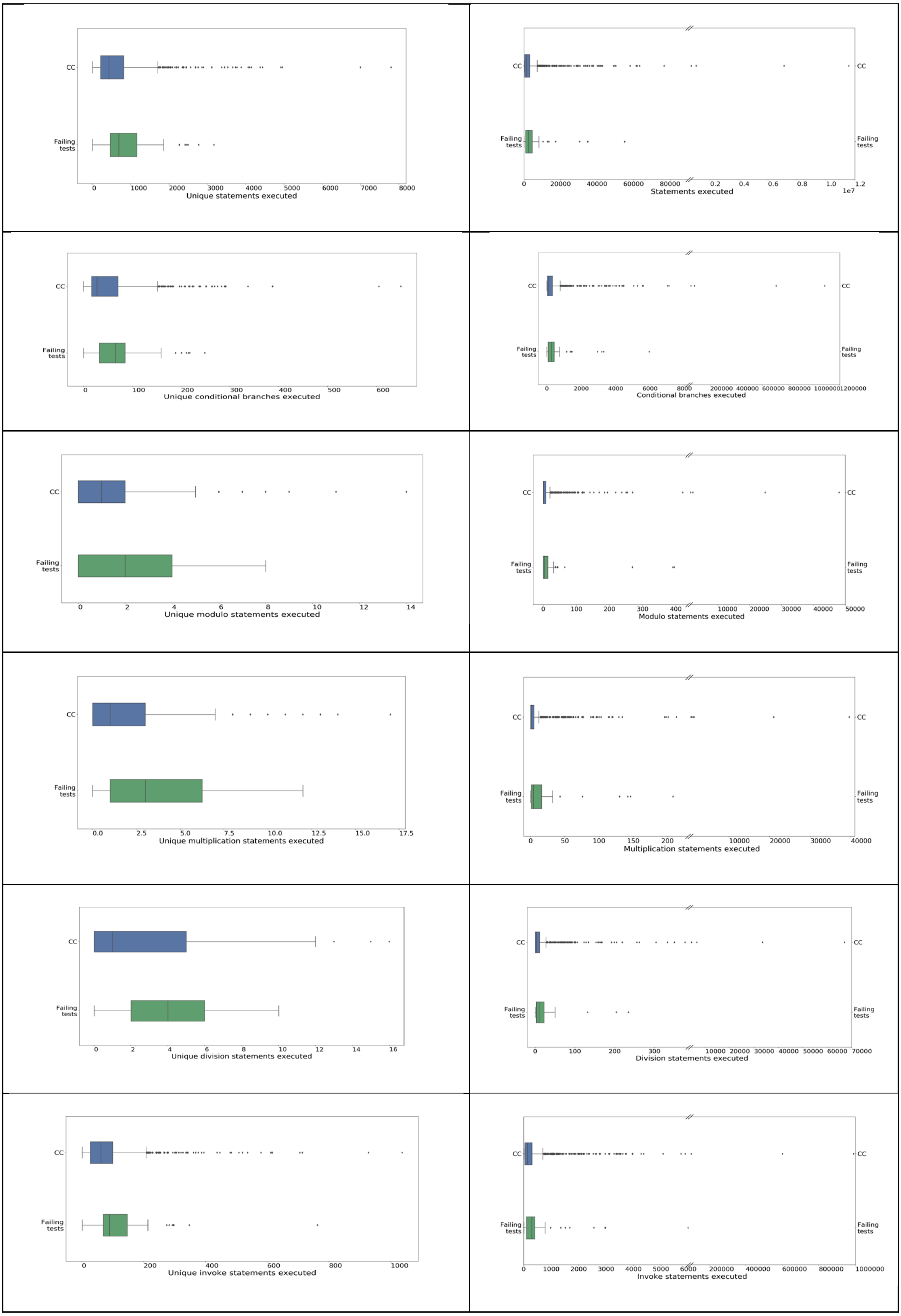

Propagation analysis for *Time*

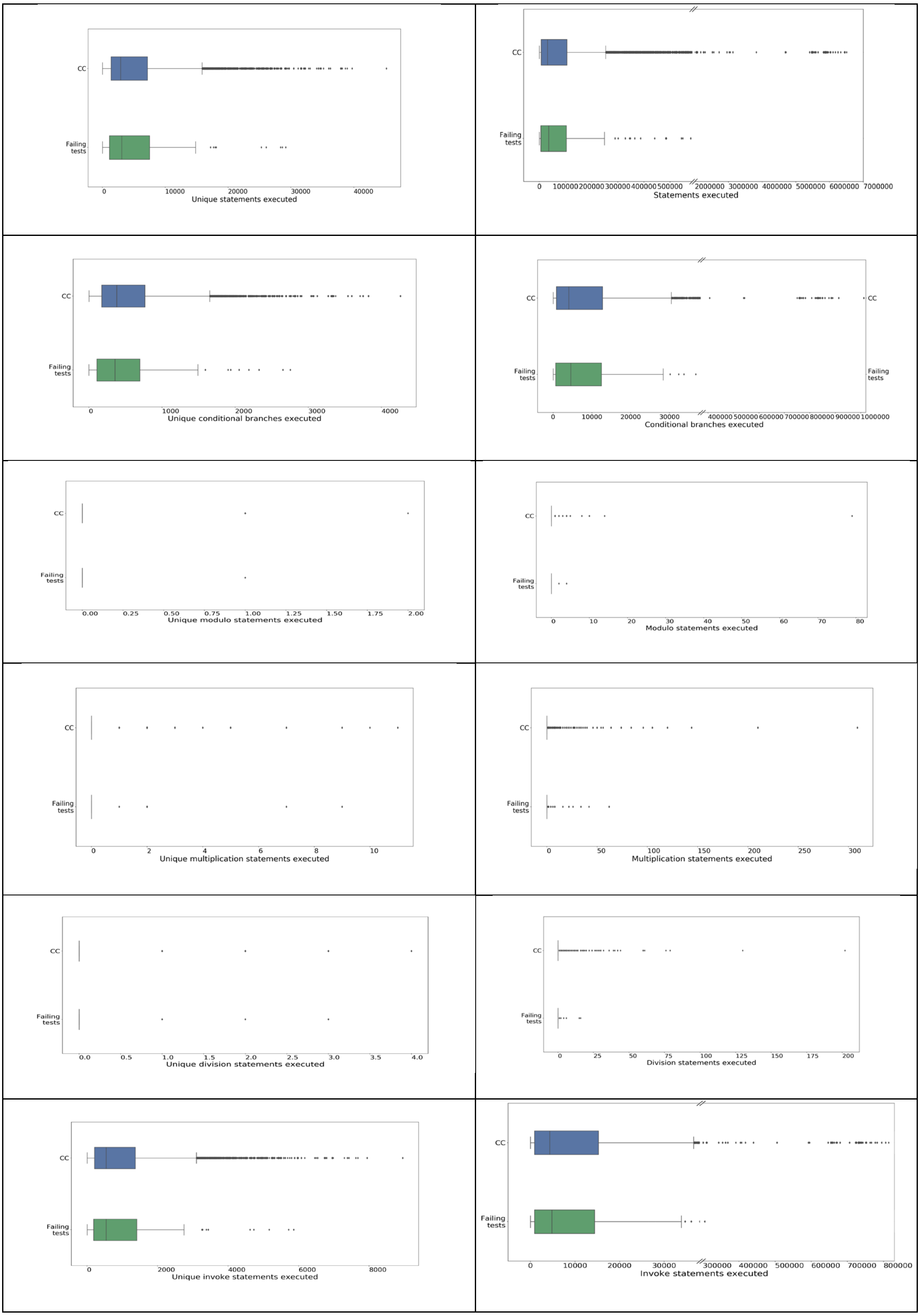

Propagation analysis for *Closure*